\pdfoutput=1
\documentclass{cta-author}

\usepackage{amsmath}
\usepackage{amssymb}
\usepackage{algorithm}  
\usepackage{algorithmicx}  
\usepackage{algpseudocode}  

\DeclareMathOperator*{\argmin}{argmin}
\usepackage{array}
\usepackage{silence}
\usepackage{subfig}
\usepackage{booktabs}
\usepackage{bm}

{}
{}
{}

\begin{document}

\title{Learning an Adaptive Model for Extreme Low-light Raw Image Processing}

\author{\au{Qingxu Fu$^{1}$}, \au{Xiaoguang Di$^{1\corr}$}, \au{Yu Zhang$^{2}$}}

\address{\add{1}{Control and Simulation Center, Harbin Institute of Technology, Harbin, 150080, People’s Republic of China}
\add{2}{National Key Laboratory of Tunable Laser Technology, Harbin Institute of Technology, Harbin, 150080, People’s Republic of China}
\email{dixiaoguang@hit.edu.cn}}


\begin{abstract}
Low-light images suffer from severe noise and low illumination. 
Current deep learning models that are trained with real-world images have excellent noise reduction, 
but a ratio parameter must be chosen manually to complete the enhancement pipeline. 
In this work, we propose an adaptive low-light raw image enhancement network to avoid parameter-handcrafting and to improve image quality. 
The proposed method can be divided into two sub-models: Brightness Prediction (BP) and Exposure Shifting (ES). 
The former is designed to control the brightness of the resulting image by estimating a guideline exposure time $t_1$. 
The latter learns to approximate an exposure-shifting operator $ES$, 
converting a low-light image with real exposure time $t_0$ to a noise-free image with guideline exposure time $t_1$. 
Additionally, structural similarity (SSIM) loss and Image Enhancement Vector (IEV) are introduced to promote image quality,
and a new Campus Image Dataset (CID) is proposed to overcome the limitations of the existing datasets and to supervise the training of the proposed model.
Using the proposed model, we can achieve high-quality low-light image enhancement from a single raw image.
In quantitative tests,
it is shown that the proposed method has the lowest Noise Level Estimation (NLE) score compared with the state-of-the-art low-light algorithms,
suggesting a superior denoising performance.
Furthermore, those tests illustrate that the proposed method is able to adaptively control the global image brightness according to the content of the image scene.
Lastly, the potential application in video processing is briefly discussed.
\end{abstract}

\maketitle
\section{Introduction}\label{sec1}

Images play an irreplaceable role in industry, military and entertainment. 
As the art of light, how to obtain better images in the low-light environment has been extensively studied in the literature. 
Some advanced photography equipment can help but at an expensive cost. 
On the other hand, there are few limitations to post-process low-light images. 
However, it is challenging due to problems such as noise, color distortion et al.

Exposure time, ISO (which measures the sensitivity of the image sensor) and aperture are known as three pillars of photography.
With extended exposure time, the low-light problem can be easily solved, 
but it is not realistic because if the camera is not fixed or the scene contains moving objects. 
Higher ISO introduces more noise and is not always available for mobile devices. As a flexible solution, 
low-light image enhancement provides an alternative for imaging in the low-light environment.

In general, current low-light image enhancement method can be classified as follows:

Classic low-light image enhancement methods, with no application of convolutional neural networks. 
Histogram equalization (HE)\cite{pizer1987adaptive} and gamma correction\cite{huang2012efficient} provide simple solution for low-light image processing. 
Based on Retinex theory\cite{land1964retinex}, Single-scale Retinex (SSR)\cite{557356}, Multi-scale Retinex (MSR)\cite{jobson1997multiscale}, 
SRIE\cite{fu2016fusion} and LIME\cite{guo2017lime} were proposed, enhancing low-light image by estimating the illumination map and 
reflectance map. Dehazing\cite{dong2011fast} methods can also be applied to this subject, regarding low-light images as 
inversed hazed image. 

With the rapid development of the deep neural network, researchers have combined classic low-light 
image enhancement methods with Convolutional Neural Networks (CNNs). Chen et al. proposed Retinex-Net\cite{wei2018deep} to decompose low-light image 
into illumination and reflectance and use BM3D\cite{Kostadin2007Image} for denoising.

Those methods assume low-light images are free of noise or noise is already removed by additional denoising process. 
Therefore, when applying those methods on real-world low-light images, 
an additional denoising process is required. 
However, most low-light images suffer from severe noise, 
traditional denoising methods like BM3D are not able to provide satisfactory results. 
To be rid of image noise, researchers also explored the method of deep learning\cite{burger2012image,lore2017llnet,zhang2017beyond,brooks2019unprocessing} in denoising topics. 

\begin{figure}[t]
  \centering
  \includegraphics[width=3in]{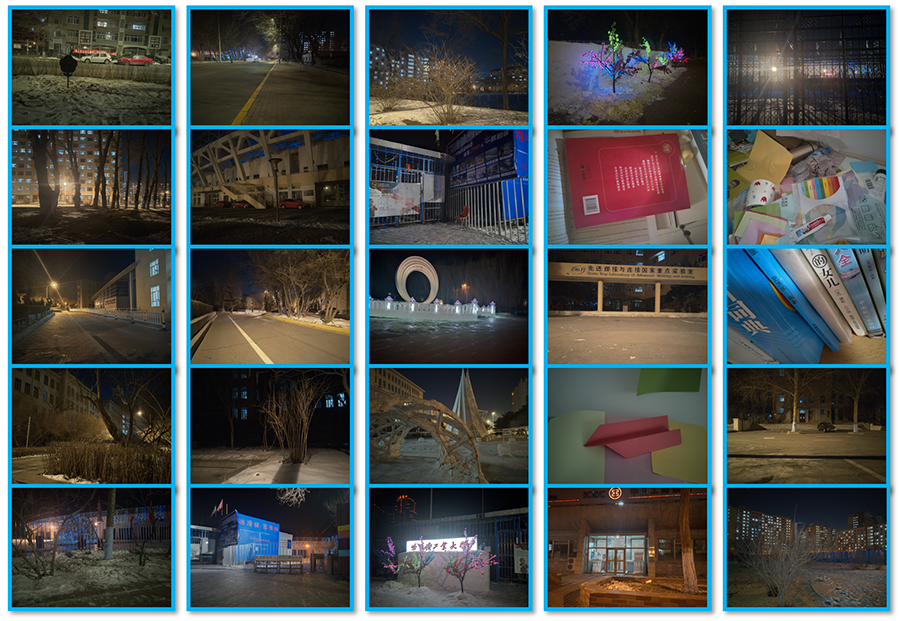}
  \caption{Several scenes in CID datasets.}
  \label{CID}
\end{figure}

There are some low-light images which are barely visible before post processing, 
as they are captured in extremely dark environments or with very short exposure time.
In this work, those images are named as extreme low-light images.
While the classic methods are unable to tackle the severe noise and serious color distortion in those images,
it is then found that with the help of deep learning and raw-format image datasets,
we can still obtain satisfactory results.
Chen et al. proposed an end-to-end solution for raw low-light image enhancement\cite{chen2018learning}, 
dealing with the low-light problem and denoising in a single model. 
However, extra brightness amplification ratio is needed and requires manual parameter tuning for different scenes, 
gaining excellent denoising effects but losing the flexibility and adaptability of classic methods.

\begin{figure*}[!t]
  \centering
  \includegraphics[width=5in]{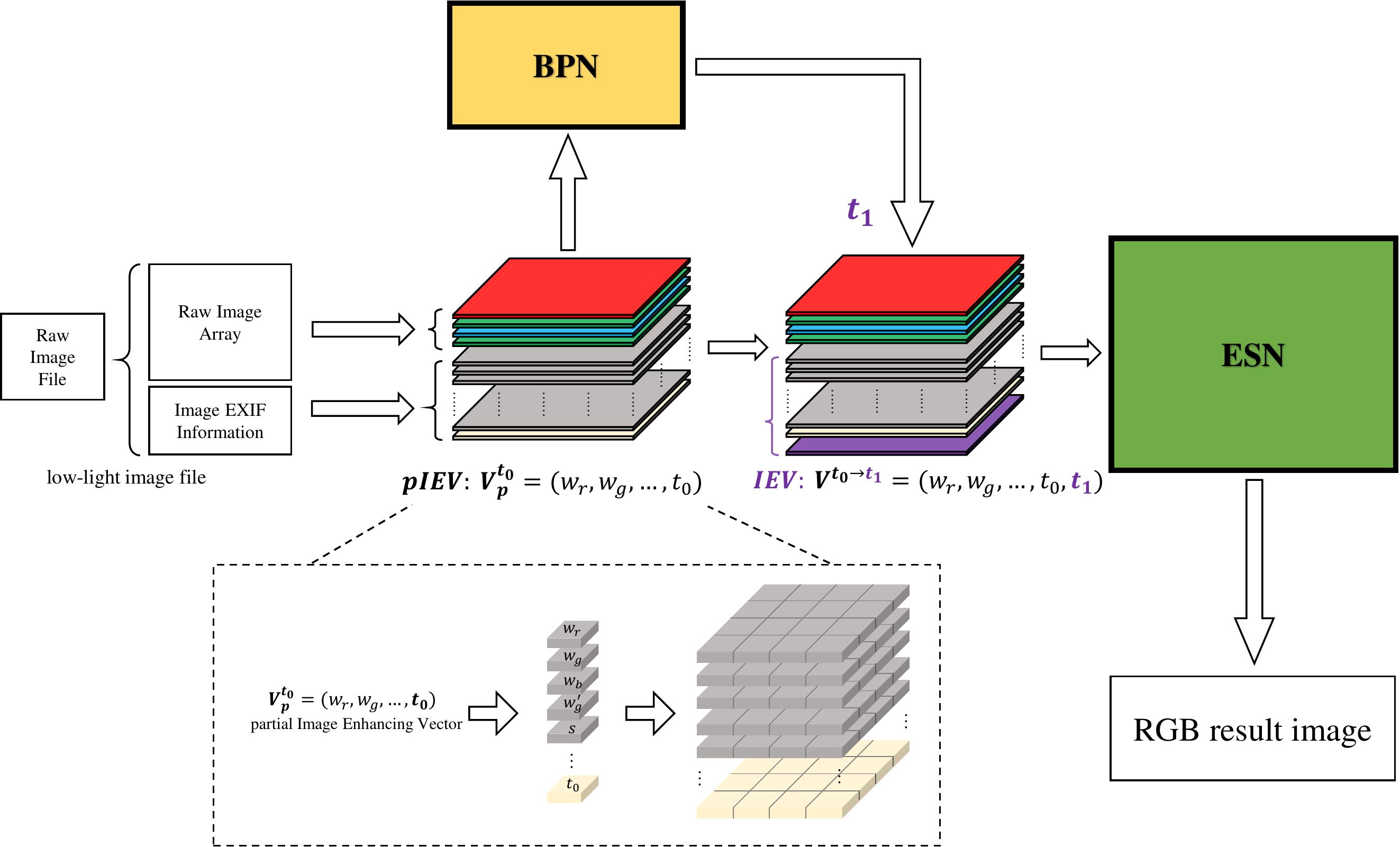}
  \caption{The proposed framework. For a underexposed low-light raw image 
  with guideline exposure time $t_1$ already known, the Exposure Shifting Network (ESN) 
  is applied to complete the estimation of the RGB result image. 
  In a more general case, Brightness Prediction Network (BPN) is 
  utilized to give a proper estimation for the uncertain 
  parameter $t_1$ with raw image array and available Exif metadata.}
  \label{fig_sim}
  \end{figure*}

Currently, all datasets available for low-light enhancement models chose a longer-exposed image as the ground-truth image in training,
which was heavily dependent on the dataset collector's experience to select exposure parameters such as exposure time, ISO and white balance mode. 
Thus these parameters were usually not optimal, considering the content of the scene, 
illumination of the environment et al., which can lead to an undesired trained model.

It is hard for deep CNNs to learn how we determine the best exposure parameters for the ground-truth images because this can be very subjective and lacks specific standards. It can lead to more problems if there is more than one dataset collector.
On the other hand, making a baseline for the ground-truth selection is equally difficult. 
Nevertheless, it is possible to predict how an image varies when exposure parameters change. 
We name this process Exposure Shifting (ES). 
With a learned ES model, it becomes possible to reversely study 
what good exposure parameters (e.g. exposure time) \textit{should} be suitable for each low-light image,
utilizing the characteristics of the back-propagation algorithm.

The contributions of our work are summarized as follows:

1) We presented Campus Image Dataset (CID), which contains a variety of scenes captured by multi-level exposure. 
CID provides powerful support for the training of our data-driven low-light enhancement model.

2) We proposed an adaptive two-stage low-light image enhancement model, which
  provides state-of-the-art low-light noise suppression as well as adaptive global brightness control effect.

In stage one, a Brightness Prediction Network (BPN) was introduced to estimate a proper exposure time based on the content of images as well as Exif (Exchangeable image file) metadata. 
BPN was trained to provide adaptive image brightness control during image enhancement
and to get rid of the external parameters required in the Exposure Shifting stage.

In stage two, an Exposure Shifting Network (ESN) was proposed to estimate a longer-exposed version of the image with target exposure time estimated by BPN. 
Moreover, ESN also completes image denoising and ends up with the final resulting RGB image.

The rest of the article is organized as follows. In Section 2, previous research works are provided. 
In Section 3, we demonstrate the importance and advantages of our CID (Campus Image Dataset) dataset. 
In Section 4, our model and its training details are provided, as well as the design concept of the model. 
The proposed model is compared with other state-of-the-art methods in Section 5, from multiple perspectives including denoising, adaptive brightness control performance and computational cost. The possible extension in video processing is briefly discussed here.
In Section 6, the advantages and the limitations of our model are concluded, and we also present the future expectations of this work.

\section{Related Works}\label{sec2}

The method we proposed was inspired not only by the application of deep neural networks,
but also by the classic methods based on the Histogram Equalization model, dehazing model and Retinex model.
\subsection{Histogram Equalization Methods}
Histogram Equalization (HE) method is used extensively to process digital images. 
The basic idea is to improve image visibility by changing the image histogram into a uniform distribution, 
which effectively increases image entropy for low-light images. 
In \cite{colorHE} it is argued that a color-invariant representation can be obtained by applying HE.
However, HE tends to cause noise amplification, 
details disappearance and color distortion in low-light image enhancement.

On the account of the drawbacks of HE, a number of improvements have been put forward. 
Adaptive Histogram Equalization (AHE) \cite{ahe} and its variation Contrast-limited Adaptive Histogram Equalization (CLAHE) \cite{CLAHE} 
perform the histogram equalization considering not only the global histogram, but also the neighbors of each pixel.
Hue-preserving color image enhancement \cite{Hue-preserving} works on contrast-enhanceing and hue-preserving,
\cite{wang1999image} and \cite{kim1997contrast} on preserving the original image luminance.

\subsection{Dehazing and Retinex Model-Based Methods}

A number of low-light enhancement methods are based on the dehazing and Retinex model.
It is observed that low-light images and inversed haze images share many similarities. 
For this reason, Dong et al. \cite{dong2011fast} presented a method and achieved the low-light image enhancement 
in which the low-light images is inversed, dehazed and then inversed back again.
This method was studied further by \cite{li2015a} and \cite{malm2007adaptive}.

On the other hand, the Retinex model \cite{land1964retinex} revealed that a natural image is made up of the illumination map and the reflectance map. 
With Retinex decomposition and the assumption that the illumination map is smooth, 
researchers have proposed single-scale Retinex \cite{jobson1997properties} and multi-scale Retinex \cite{557356}. 
Based on image fusion, adjustments can be applied to the illumination map to improve the performance \cite{un24}.
\cite{un23} focused on preserving natural characteristics with a Bright-Pass filter.
In \cite{kimmel2003a}, a variational Retinex model was proposed and the illumination estimation problem was formulated as a Quadratic Programming optimization problem. 
In \cite{refinedretinex} illumination map is estimated considering local structure 
and \cite{retinexnoisemap} focused on revealing the structure details from the reflectance map.
A recent study shows that the Retinex model can combine with the atmospheric scattering model\cite{retinexscatter}.
More refined models were presented by \cite{un28,un29} based on variational Retinex theory.

However, noise modeling is not well established in those algorithms.
In \cite{dong2011fast}
it is assumed that noise is insignificant and can be removed before or after the Dehazing stage. 
And the Retinex based models rely on classic noise reduction algorithms (e.g. BM3D) to reduce noise found in the reflectance map. 
Therefore it can achieve good results in mild low-light images
but tends to amplify noise when processing severe or extreme low-light images. 
Besides, the variational Retinex model is very time-consuming as it takes many iterations to solve the optimization objectives, 
thus it is hard to achieve real-time processing.

\subsection{Data-driven Methods}

Image-denoising and low-light enhancement can be realized with the data-driven approach, separately or integrally.

Noise reduction was the subject of extensive studies. 
Burger et al.\cite{burger2012image} discovered that convolutional neural networks could compete with the state-of-the-art classic denoising algorithm BM3D. 
The works on date-driven denoising have been extended by \cite{un31,un32,brooks2019unprocessing}. 
Although these works were not targeted at low-light image processing, 
they provided some references for the research of other image processing studies.

The first application of the data-driven method is LLNET \cite{lore2017llnet}, 
which utilized a deep auto-encoder to learn from synthetic low-light image datasets and the denoising procedure is integrated into the low-light enhancement process.
Also, it was demonstrated that Retinex based methods can be further improved by deep learning \cite{wei2018deep,shen2017msr}, 
but those works still have undesirable denoising performance because synthetic datasets are used and they cannot reflect the characteristics of real low-light images.

To solve the drawbacks of synthetic datasets, a large multi-exposure image dataset was proposed in \cite{cai2018learning}.
It was found by Chen et al.\cite{chen2018learning} that models utilizing raw-format images can provide much better results in low-light enhancement.
In the SID model proposed in \cite{chen2018learning}, 
it suggested that the state-of-the-art results can be achieved by using raw-format paired images and training in an end-to-end way. 
However, since the paired images have different exposure time,
a ratio was introduced as an additional input to bridge the gap.
Consequently, if the SID model is applied on images out of the dataset, 
manual adjustment is requires because there is no paired images to indicate the ratio,
which results in many drawbacks in practical applications.

\section{Dataset}\label{sec3}

For extreme low-light image enhancement, there are few optional datasets.
The Google HDR+ dataset\cite{hasinoff2016burst} and Darmstadt Noise Dataset\cite{xu2018real} avoid the disadvantages of synthetic datasets, but they were mainly collected in environments with sufficient illumination.
The RENOIR\cite{anaya2014renoir} and learning-to-See-In-the-Dark dataset (SID)\cite{chen2018learning} provides real low-light image pairs, 
which is captured by carefully selecting the ISO and exposure time for each scene.
The Exclusively Dark Dataset (EDD) \cite{Exdark} is another low-light dataset with object-level annotations, which is targeting at object recognition.
But these datasets cannot meet the requirements of our study.
In our work, in order to adaptively enhance images with different noise levels, 
it is needed to consider how the exposure parameters have an impact on low-light images.
More specifically, as the environment illumination grows lower, low-light images are gradually overwhelmed by noise and invalid pixels.
Although it is not flexible to manipulate the environment illumination, 
the camera exposure time setting can be easily changed to simulate this illumination shift.
Therefore, it is expected that there is a dataset containing, instead of just some image pairs, a number of multi-exposed image series, 
in each of which there are low-light images of different levels and one reference image captured in the same scene.

Overall, there are three conditions to be met for a satisfactory dataset:
\begin{itemize}
  \item Real scenes. 
Unlike synthetic images, 
photos captured in real scenes reflect much more diverse noise and distortion pattern

  \item Multiple exposure levels. 
Using a number of multi-exposed image series to train an adaptive model 
capable of enhancing low-light images of different levels.

  \item Raw-format images. 
Most data captured by the camera sensor is lost in format convertion, 
those lost infomation is essential for extreme low-light image enhancement.
\end{itemize}
\begin{figure}[htb]
  \centering
  \includegraphics[width=3.4in]{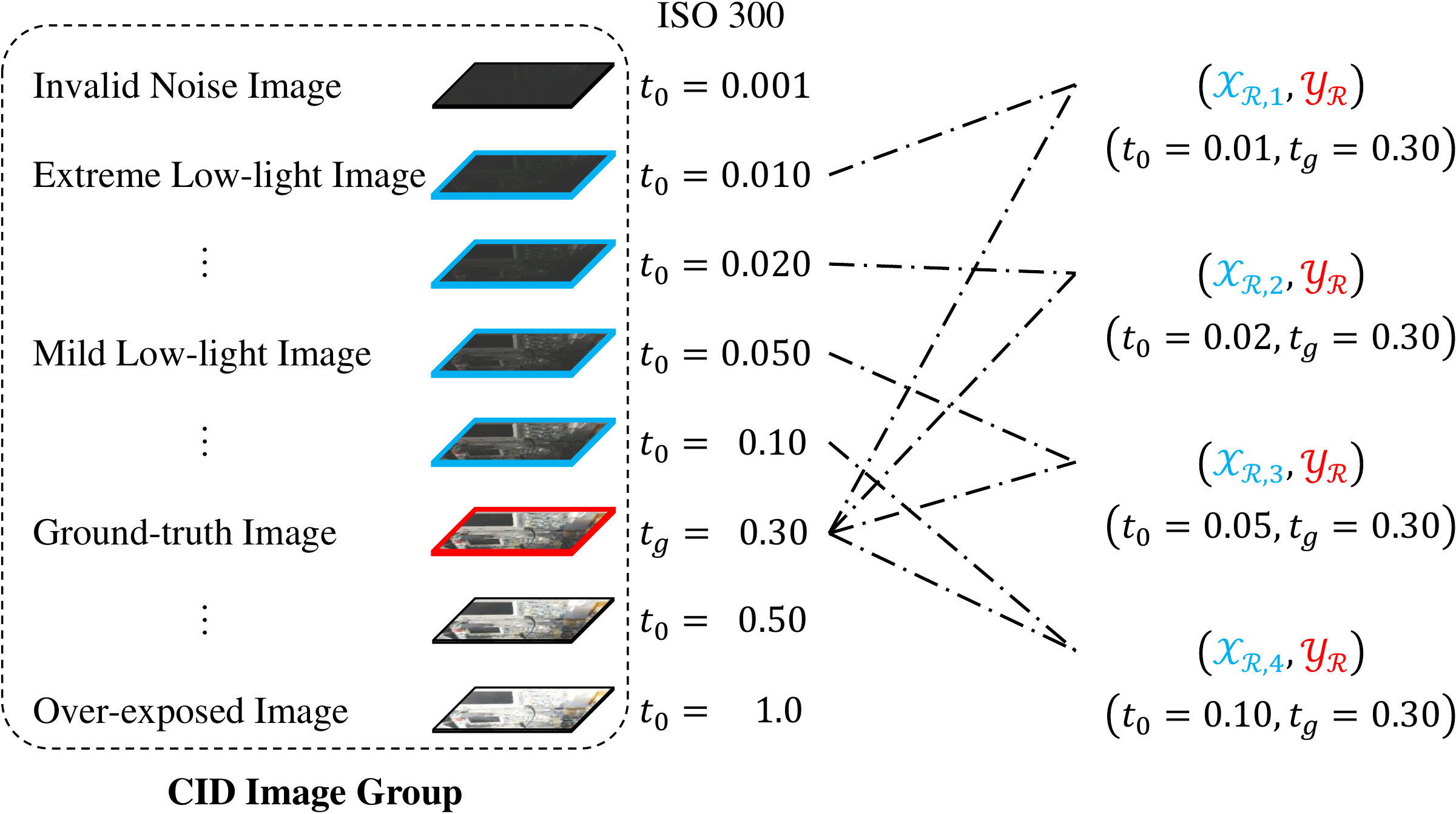}
  \caption{Demonstration of a CID image group.}
  \label{cidg}
  \end{figure}
Based on the above description, we proposed our Campus Image Dataset (CID)\footnote{github.com/505030475/ExtremeLowLight}. 
There are about 200 image groups in CID, each of which
consists of 8 raw images with different exposure time, shot continuously in the same scene using a tripod (Fig.\ref{cidg}). 
To be specific, for every group, an upper limit of exposure time was selected to make sure that the first image in the sequence was slightly overexposed, 
then a lower limit was chosen to capture an extreme low-light image as the last image in the sequence, 
and the gap between the limits was filled up by another 6 images. 
Finally, the ground-truth image for each group was manually selected, 
and invalid images that had nothing but noise due to extreme short exposure were tagged and excluded.
A demonstration of scenes in CID is presented in Fig.\ref{CID}.

\section{Method}\label{sec4}

\subsection{Method Overview}

Most of the existing methods are unable to suppress the severe noise present in extreme low-light images. 
On the other hand, 
recent studies e.g. \cite{chen2018learning} have concentrated on the extreme low-light enhancement with raw-format images and real-scene datasets in an end-to-end way,
but unlike the classic methods, 
the brightness of the enhanced image cannot be automatically controlled in this framework, 
more specifically, 
an external ratio has to be manually adjusted when the input image has no paired image as the reference.

It is our aim to find a solution to enhance extreme low-light images, 
combining the advantages of adaptive brightness control and superior denoising performance.
It is found that the problem of low-light enhancement can be considered as a special case of changing the time of exposure.
With $t_0$ and $t_1$ representing exposure time,
the low-light image $\mathcal{X}_{t_0}$ can be interpreted as a normal image $\mathcal{Y}_{t_1}$ corrupted by a hypothetical exposure-shifting operator $ES$,
which only alters the exposure time but keeps the aperture and ISO unchanged:
\begin{equation}B
  \label{L-1}
  \mathcal{X}_{t_0} = ES(\mathcal{Y}_{t_1},t_0)
\end{equation}
In this process, 
more stochastic noise $\mathcal{N}$ is generated by operator $ES$ because $t_0<t_1$, 
corresponding to the low PSNR (Peak Signal to Noise Ratio) value in low-light images.
Contrariwise, the corrupted low-light image can also be restored by the same operator $ES$: 
\begin{equation}
  \label{L-2}
  \hat{\mathcal{Y}}_{t_1} = ES(\mathcal{X}_{t_0},t_1)
\end{equation}
The noise $\mathcal{N}$ is reduced because $t_1>t_0$. Thus the normal image $\hat{\mathcal{Y}}_{t_1}$ is reconstructed.

The latter process, named as Exposure Shifting (ES), can enhance low-light images as well as suppress image noise.
More importantly, it can be learned in a data-driven and end-to-end way, thus it is expected to have superior denoising performance.
However, a proper guideline exposure time $t_1$ always has to be provided in the reconstruction $\hat{\mathcal{Y}}_{t_1}$.
To make our model adaptive, a Brightness Prediction (BP) procedure is introduced to achieve guideline time estimation.

In brief, the extreme low-light enhancement process is split into two procedures.
Firstly, for a raw-format low-light image (with real exposure time $t_0$), 
an optimal guideline exposure time $t_1$ is estimated via a sub-model called Brightness Prediction Network (BPN).
Secondly, the final RGB-format enhanced image is obtained by another sub-model, namely Exposure Shifting Network (ESN), 
using estimated guideline time $t_1$ to control the image brightness.

In addition, besides the image itself, some extra data are involved in both BPN and ESN, 
such as white balance index, ISO,  $t_0$ and $t_1$. 
The Image Enhancing Vector (IEV) is put forward to encode the involved extra data.

\subsection{Image Enhancing Vector}

Raw image array is the original data captured by the camera. Other key parameters, such as camera white balance, 
ISO, exposure time, time and date, are stored together with the raw image array as Exif metadata. 
In HDR imaging\cite{Reinhard2005High}, 
exposure time and ISO information have played an essential role in calculating the response curve and LDR to HDR conversion. 
Although it is not necessary to explicitly train the network to learn the camera response curve, 
feeding these additional parameters into the network as input avoids underfitting caused by insufficient features.

Image Enhancing Vector (IEV) is proposed to introduce extra features into the convolution neural networks. 
In this paper, IEV consists of ISO $u_{s}$,
white balance indices for 4 raw image channels $(w_r,w_g,w_b,w_{g2})$, exposure time of the 
low-light image $t_0$ and guideline exposure time $t_1$. 
Moreover, it is possible to append other Exif metadata into IEV to improve 
network performance when necessary,e.g. aperture and focal length.

\begin{figure}[!t]
  \centering
  \includegraphics[width=3in]{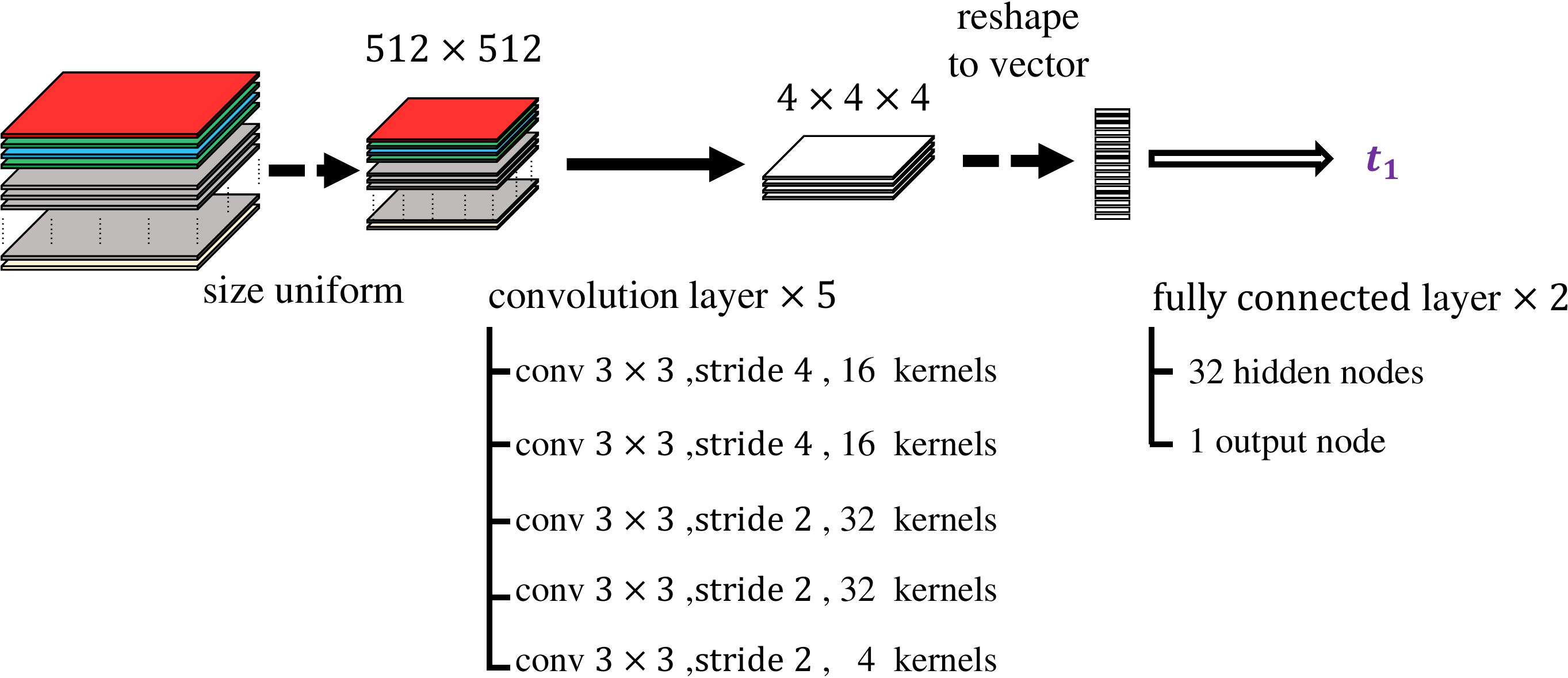}
  \caption{The components of BPN.}
  \label{bpn}
  \end{figure}

The IEV is used as network input in both BPN (for $t_1$ estimation) and ESN (for exposure shifting),
but a difference exists. 
As it is shown in Fig.\ref{fig_sim}, in ES procedure the IEV can be written as:
\begin{equation}
  \label{L-iev}
  \mathcal{V}^{t_0 \rightarrow t_1} = \mathcal{V}(t_0,t_1) = (w_r,w_g,w_b,w_{g2},u_{s},...,t_0,t_1)
\end{equation}
Where $(w_r,w_g,w_b,w_{g2})$ are the white balace indices of 4 raw image channels and $u_{s}$ is the ISO value (controlling the camera sensitivity).

In BP procedure the $t_1$ in IEV is removed. It is renamed as partial IEV (pIEV) to indicate the difference:
\begin{equation}
  \label{L-piev}
  \mathcal{V}_p^{t_0} = \mathcal{V}_p(t_0) = (w_r,w_g,w_b,w_{g2},u_{s},...,t_0)
\end{equation}

IEV and pIEV are fed into the convolutional neural network as additional channels of the raw image array.
For example, the first extra channel provided by IEV is a uniform image filled up with pixel value $w_r$.

\subsection{Exposure Shifting}

Exposure Shifting (ES) is the second procedure in our model but is trained first.
The U-NET\cite{ronneberger2015u} is selected as the structure of the Exposure Shifting Network (ESN),
Compared with the fully convolutional network, the U-NET architecture reduced the usage of GPU memory, 
thus it can easily process images with high resolution.

Recall that the basic unit of our Campus Image Dataset (CID) is an image group,
which consists of 8 raw-format images with different exposure time but captured in the same scene.
Let $\left(\mathcal{X}_R,\mathcal{Y}_R\right)$ be the paired raw image extracted from one group, 
in which $\mathcal{Y}_R$ is the pre-selected ground-truth image 
and $\mathcal{X}_R$ is a randomly-selected low-light image.
More specifically, $\mathcal{X}_R$ is randomly chosen within this image group, 
with $\mathcal{Y}_R$ and over-exposed images excluded (Fig.\ref{cidg}).
Moreover, the exposure time of $\mathcal{X}_R$ is denoted as $t_0$.
The corresponding RGB-format image of $\mathcal{Y}_R$ is denoted as $\mathcal{Y}$ and its exposure time as $t_g$.

It is expected for the ESN to learn the exposure-shifting operator $ES$ from a large number of paired images.
Specifically, 
the ESN is required to estimate an image $\hat{\mathcal{Y}}$ that resembles the $\mathcal{Y}$ as closely as possible.
In this way ESN successfully changes the raw-format low-light image $\mathcal{X}_R$ into an RGB-format longer-exposed version of itself. 
More importantly, the serious noise of $\mathcal{X}_R$ can also be removed. This procedure can be written as:
\begin{equation}
\begin{aligned}
  \label{L_esn}
  &\hat{\mathcal{Y}}^{\Theta_1}_{t_g}=F_{ES}\left(\mathcal{X}_{R},\mathcal{V}^{t_0 \rightarrow t_g} \middle| \Theta_1\right)  \\
  &\mathcal{V}^{t_0 \rightarrow t_g} = \mathcal{V}(t_0, t_1 = t_g)
\end{aligned}
\end{equation}
Where ESN is denoted as $F_{ES}$, its parameters as $\Theta_1$. 
Note that since the BP procedure has not yet involved, the guideline exposure time $t_1$ is replace by $t_g$ in IEV $\mathcal{V}^{t_0 \rightarrow t_g}$.
The enhanced result image is represented by $\hat{\mathcal{Y}}^{\Theta_1}_{t_g}$.

Two metrics to guide the training process of the ESN model are Mean Square Error (MAE) and multi-scale Structural Similarity (SSIM)\cite{wang2003multiscale}. 
The former is considered as a simple but effective indicator, evaluating the general similarity of the estimated image and the groud-truth image.
And the later is a perceptual metric that is more sensitive to visible structures. 
Both of them are full-reference metrics.
Let $L_{MAE}$ be the MAE loss and $L_{SSIM}$ be the SSIM loss. 
The width and length of the image are denoted as $m$ and $n$ respectively.
With subscript $t_g$ omitted, the MAE loss $L_{MAE}$ and SSIM loss $L_{SSIM}$ are defined as:
\begin{equation}
\begin{aligned}
  \label{L2}
  &L_{MAE}\left(\hat{\mathcal{Y}}^{\Theta_1},\mathcal{Y}\right)=MAE\left(\hat{\mathcal{Y}}^{\Theta_1},\mathcal{Y}\right) 
    = \frac{1}{mn}\sum_{i=1}^{m}\sum_{j=1}^{n} \left|\hat{\mathcal{Y}}_{ij}^{\Theta_1} - \mathcal{Y}_{ij}\right|    \\
  &L_{SSIM}\left(\hat{\mathcal{Y}}^{\Theta_1},\mathcal{Y}\right)=1-SSIM\left(\hat{\mathcal{Y}}^{\Theta_1},\mathcal{Y}\right)
\end{aligned}
\end{equation}
The final MAE loss is the average of the channel MAE loss, for simplicity, the image channel is not presented in the equation.
The calculation of the multi-scale structural similarity (SSIM) can be refered in \cite{wang2003multiscale}. 
With subscript $t_g$ omitted,
the loss function is defined as the linear combination of $L_{MAE}$ and $L_{SSIM}$:
\begin{equation}
  \begin{aligned}
  \label{L1}
  {L}_{ES}
      &={L}_{ES}\left(\hat{\mathcal{Y}}^{\Theta_1},\mathcal{Y} \right) \\
     &=\left(1-\alpha\right){L}_{MAE}\left(\hat{\mathcal{Y}}^{\Theta_1},\mathcal{Y}\right)+\alpha {L}_{SSIM}\left(\hat{\mathcal{Y}}^{\Theta_1},\mathcal{Y}\right)
  \end{aligned}
\end{equation}
Where $\alpha$ is a constant and $0 \leqslant \alpha < 1$. The influence of $\alpha$ will be discussed in the \textit{Experiments} section.

The parameters of the ESN network $\Theta_1$ are learned by minimizing ${L}_{ES}$ over $K$ pairs of images in the training set:
\begin{equation}
  \label{L3}
  \Theta_1^\ast=\mathop{\argmin}_{\Theta_1} {\sum_{k=1}^{K}{L_{ES} \left(  {\hat{\mathcal{Y}}}^{\Theta_1}_k , \mathcal{Y}_k    \right)    }}
\end{equation}
$\Theta_1^\ast$ represents the optimized ESN parameters. 

The obtained sub-model $F_{ES}$ is an approximation to the exposure-shifting operator $ES$.
However, for a low-light image out of the training set, a ground-truth counterpart does not exist,
thus the guideline time $t_1$ in IEV $\mathcal{V}^{t_0 \rightarrow t_1}$ is absent, 
which leads to the Brightness Prediction (BP) sub-model and guideline time estimation.

\subsection{Brightness Prediction}
\begin{figure*}
  \centering
  \begin{tabular}{ccccc}

    \rotatebox[origin=c]{90}{low-light image}
    &
    \begin{minipage}{0.2\textwidth}
    \includegraphics[width=\textwidth]{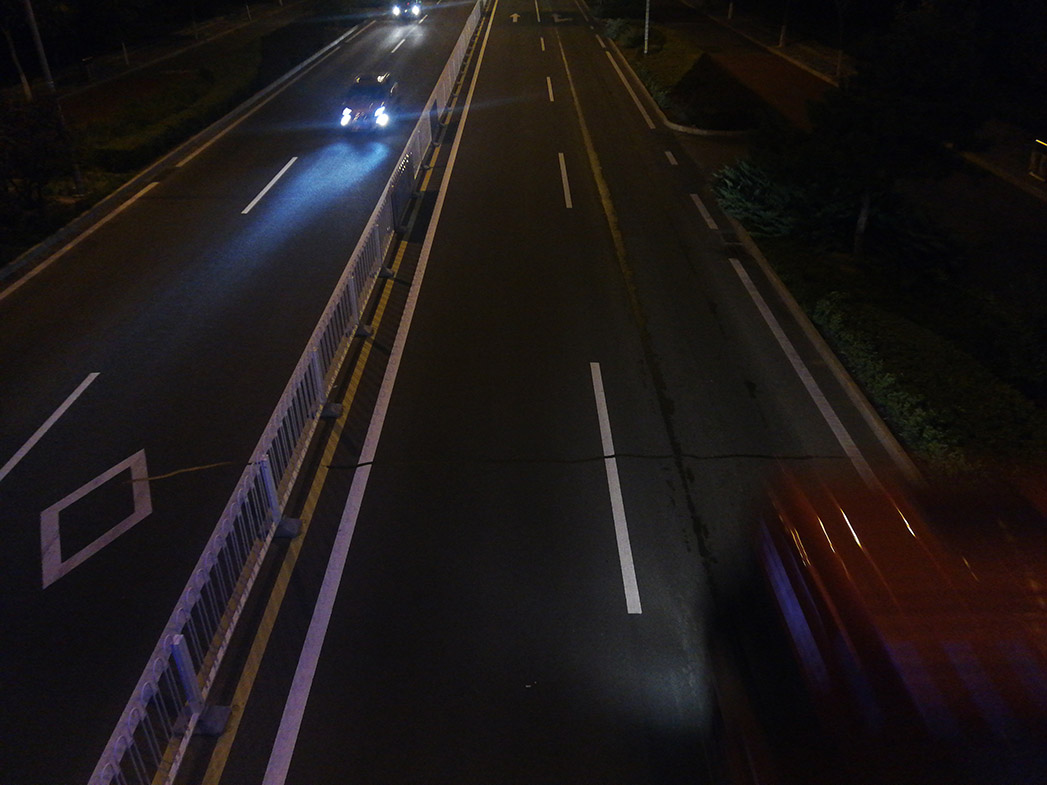}
    \end{minipage}   
    &   
    \begin{minipage}{0.2\textwidth}
    \includegraphics[width=\textwidth]{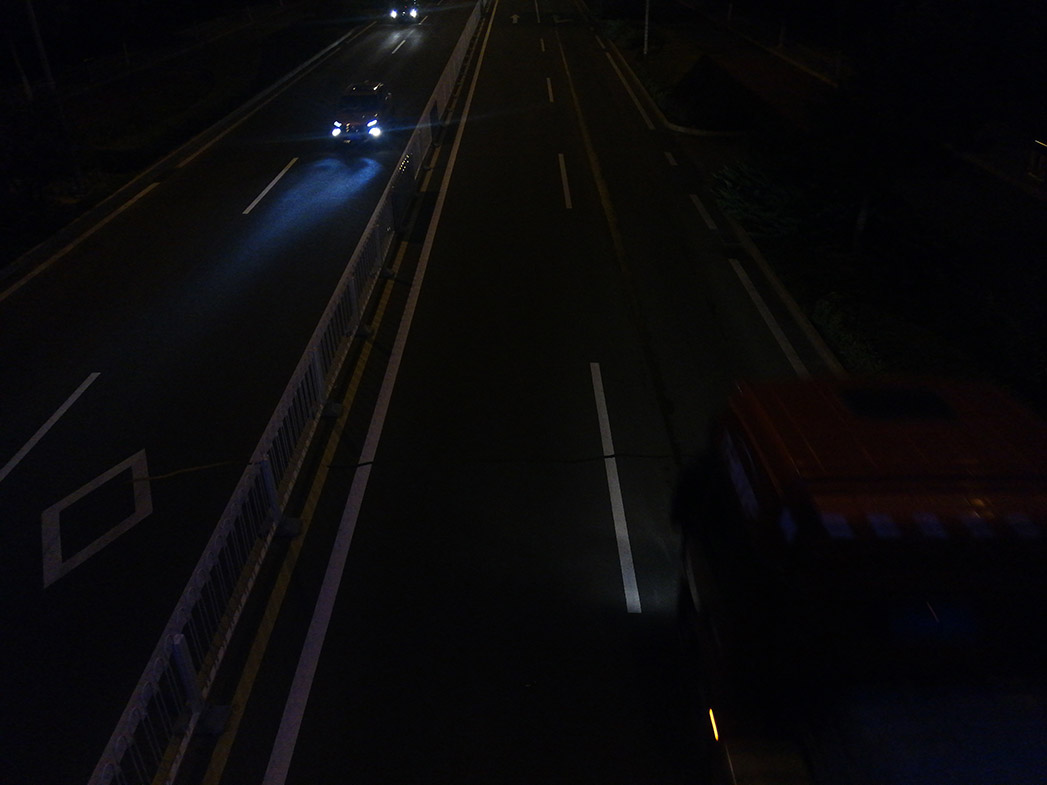}
    \end{minipage}   
    &   
    \begin{minipage}{0.2\textwidth}
    \includegraphics[width=\textwidth]{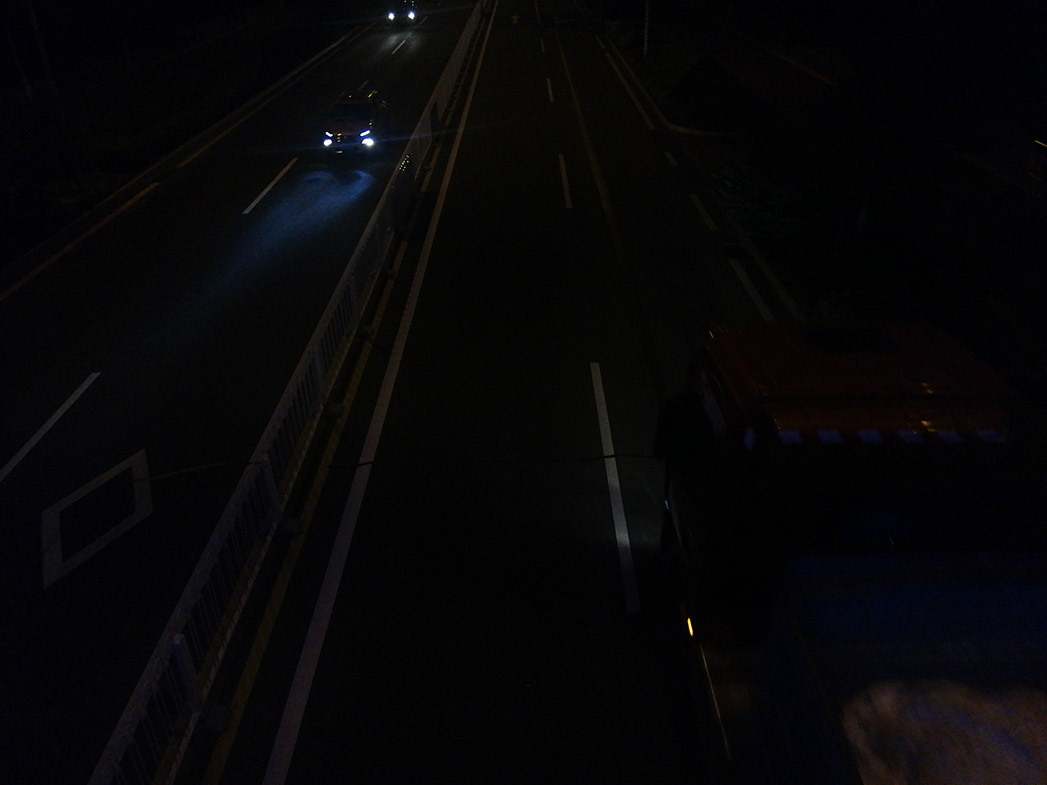}
    \end{minipage}   
    &   
    \begin{minipage}{0.2\textwidth}
      \includegraphics[width=\textwidth]{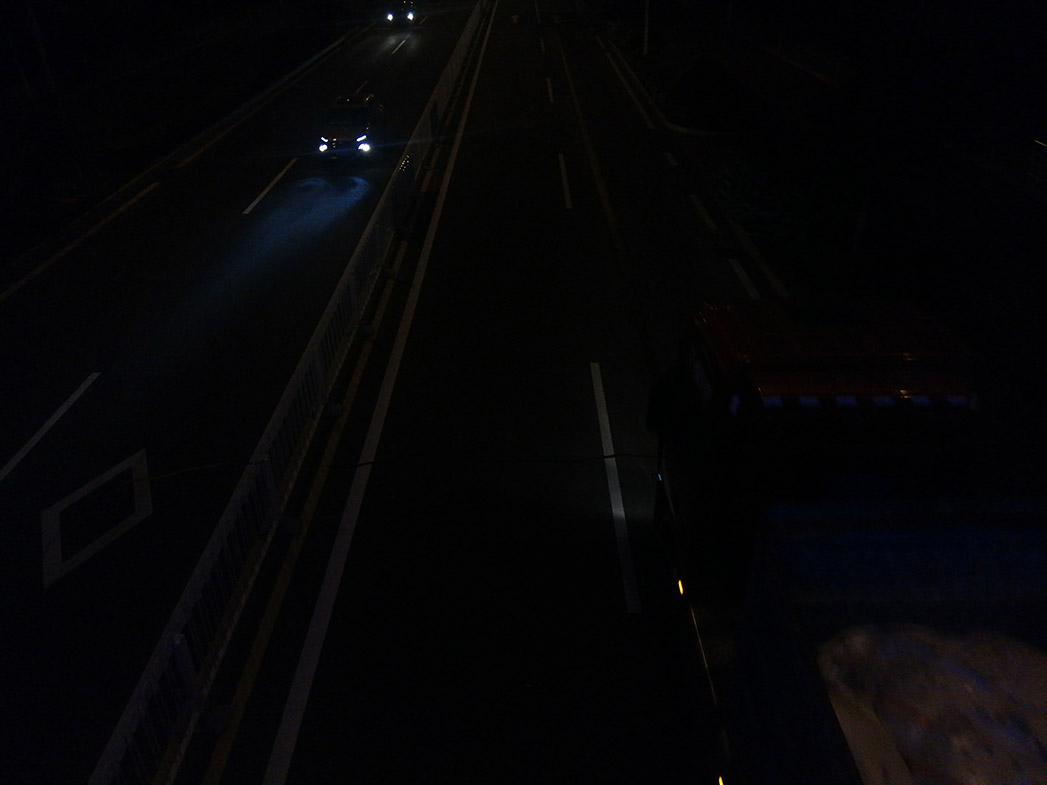}
      \end{minipage}   
     
    \\   
  
    \rotatebox[origin=c]{90}{our method}
    &
    \begin{minipage}{0.2\textwidth}
    \includegraphics[width=\textwidth]{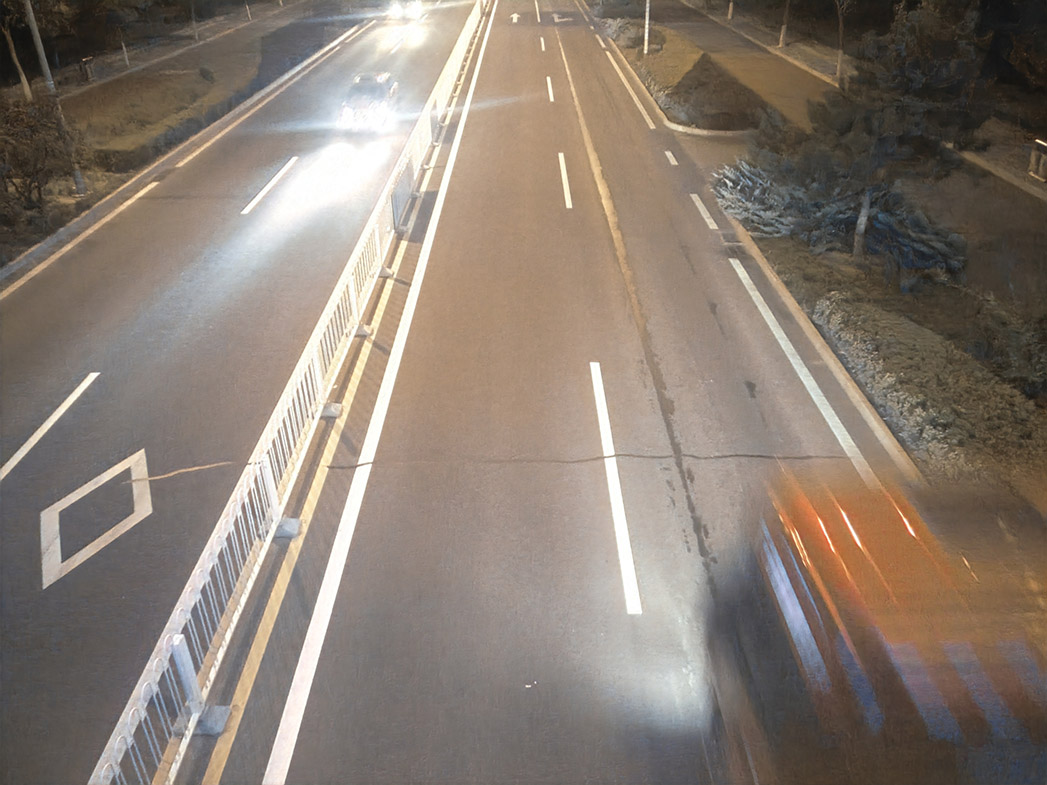}
    \end{minipage}   
    &   
    \begin{minipage}{0.2\textwidth}
    \includegraphics[width=\textwidth]{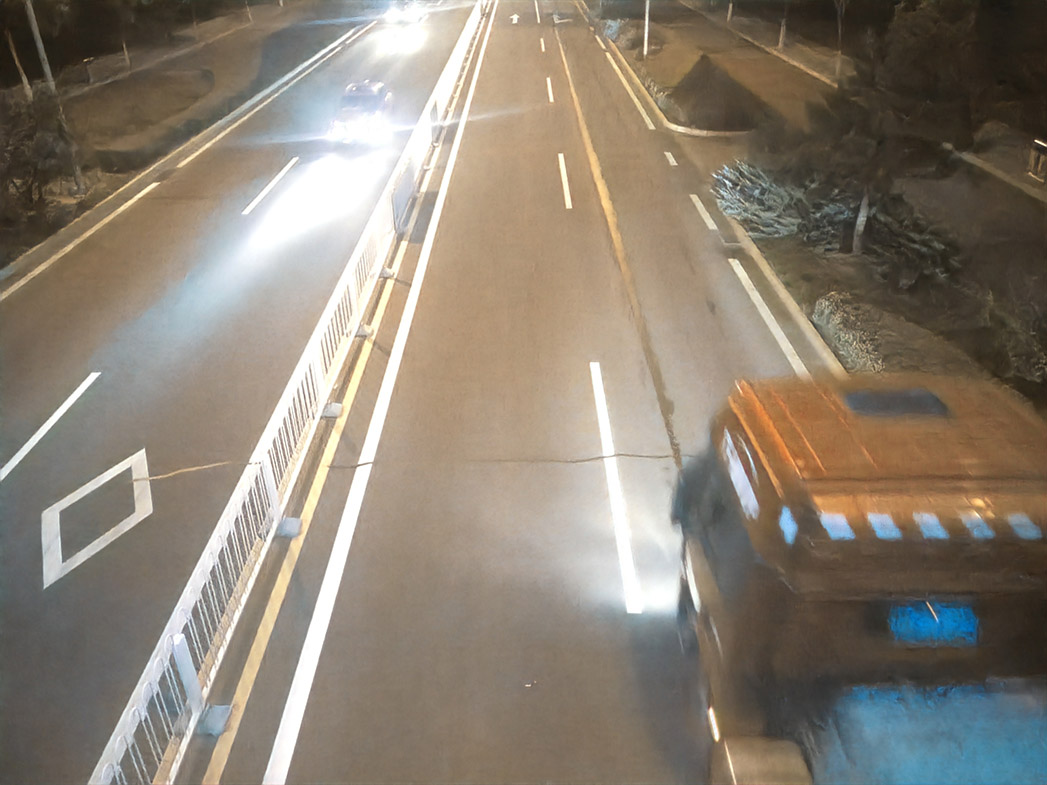}
    \end{minipage}   
    &   
    \begin{minipage}{0.2\textwidth}
    \includegraphics[width=\textwidth]{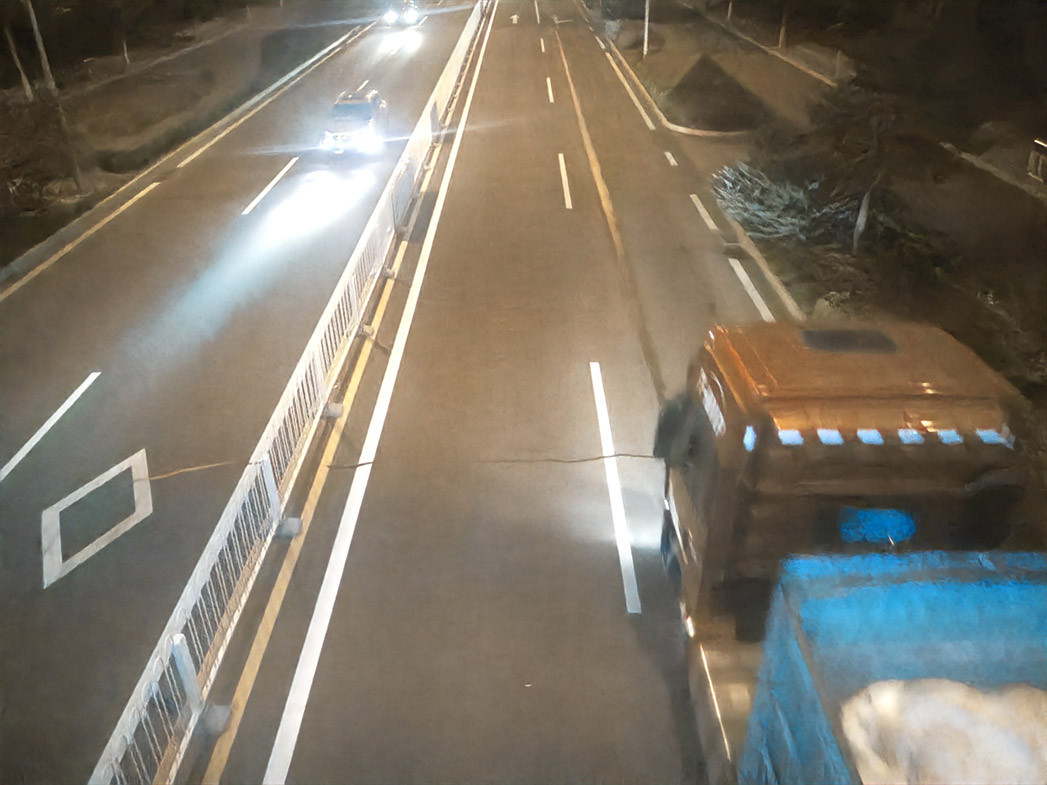}
    \end{minipage}   
    &   
    \begin{minipage}{0.2\textwidth}
    \includegraphics[width=\textwidth]{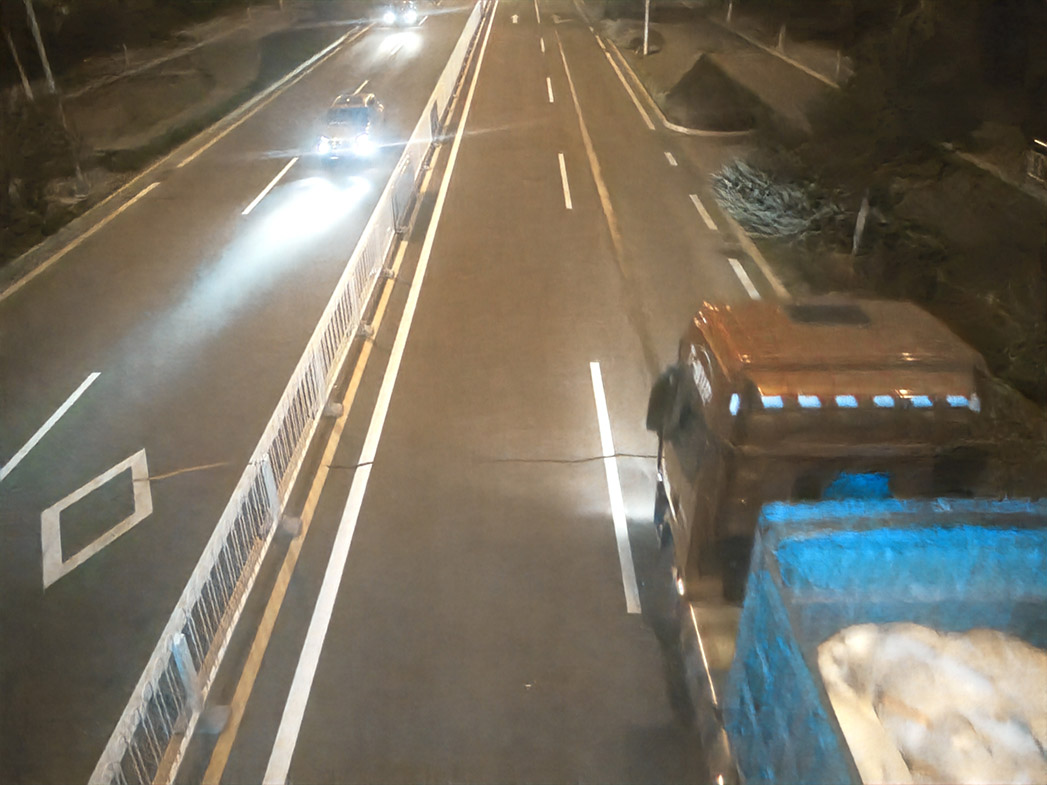}
    \end{minipage}   
    \\   

    (a) & 1/20 seconds & 1/50 seconds & 1/80 seconds  & 1/100 seconds  
     
     \\
    
    \rotatebox[origin=c]{90}{low-light image}
    &
    \begin{minipage}{0.2\textwidth}
    \includegraphics[width=\textwidth]{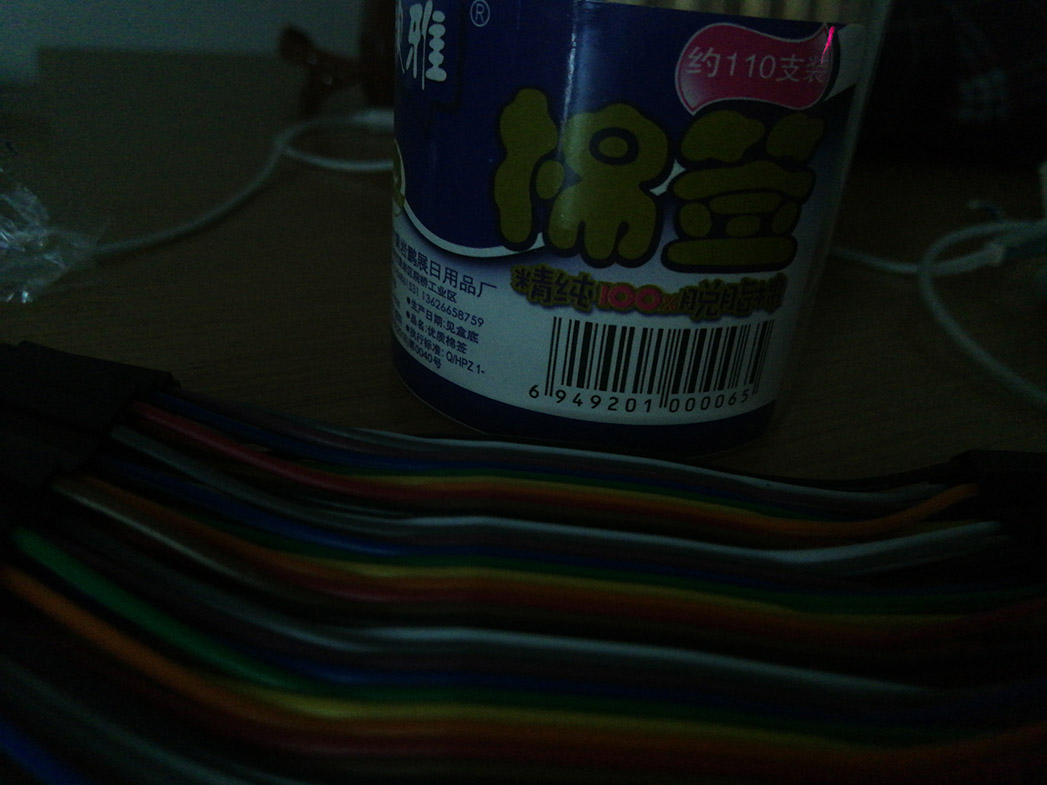}
    \end{minipage}   
    &   
    \begin{minipage}{0.2\textwidth}
    \includegraphics[width=\textwidth]{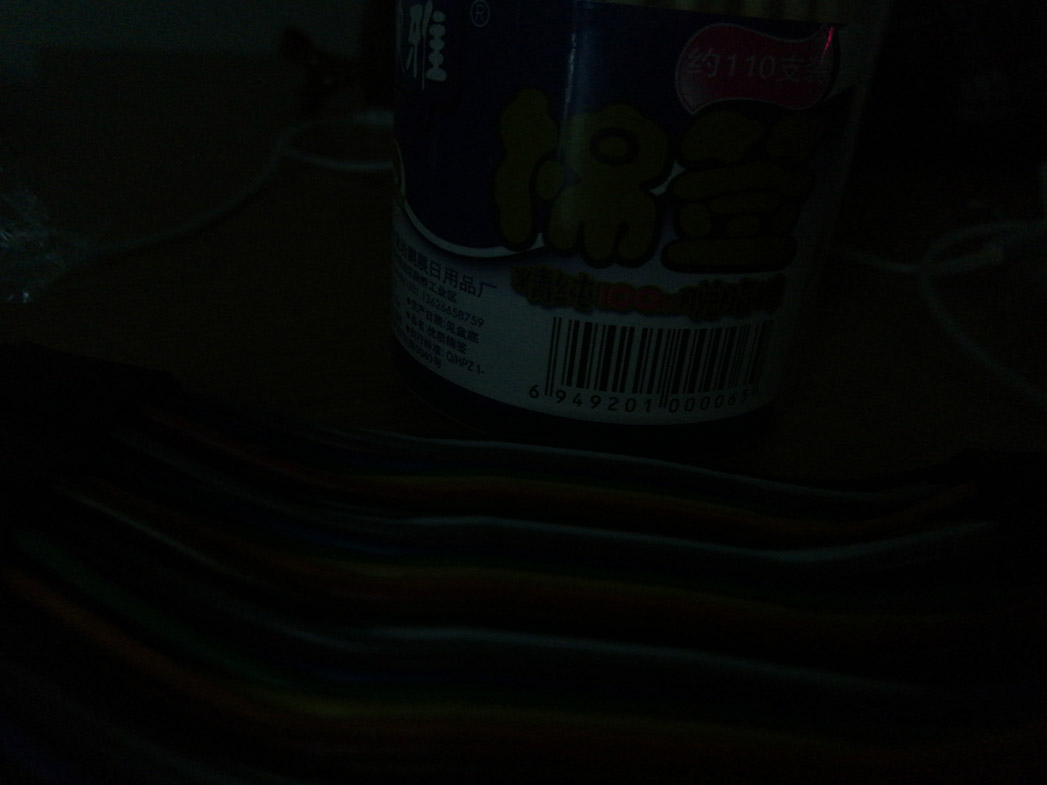}
    \end{minipage}   
    &   
    \begin{minipage}{0.2\textwidth}
    \includegraphics[width=\textwidth]{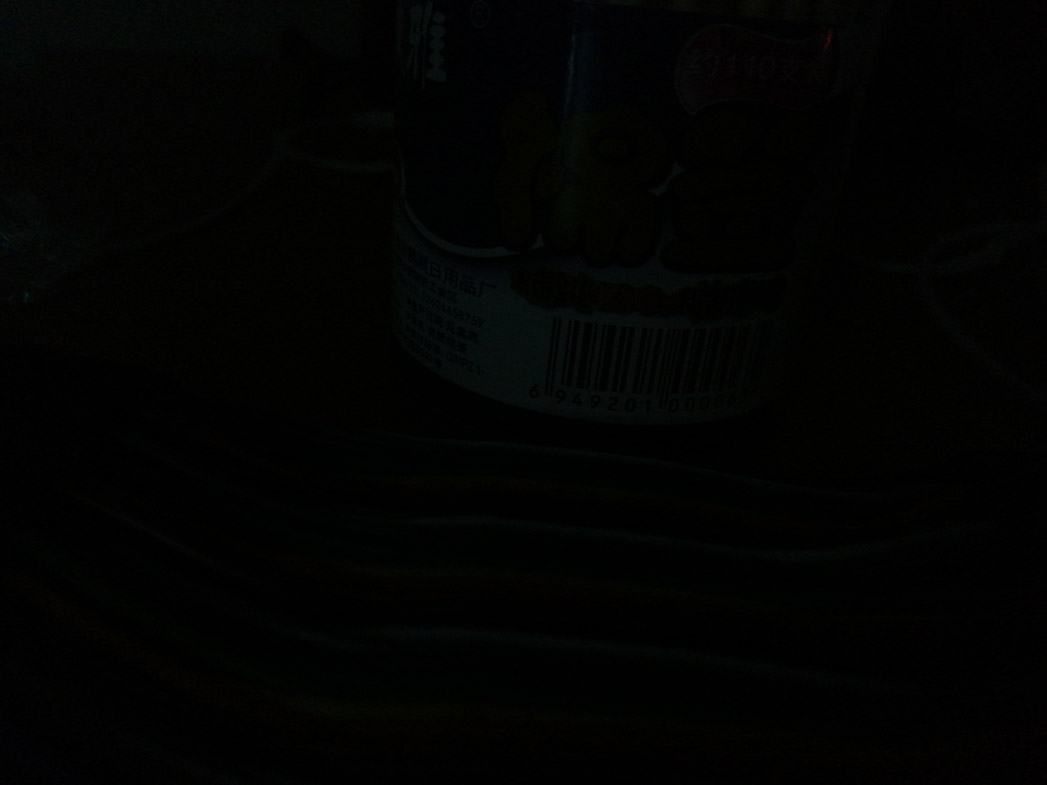}
    \end{minipage}   
    &   
    \begin{minipage}{0.2\textwidth}
    \includegraphics[width=\textwidth]{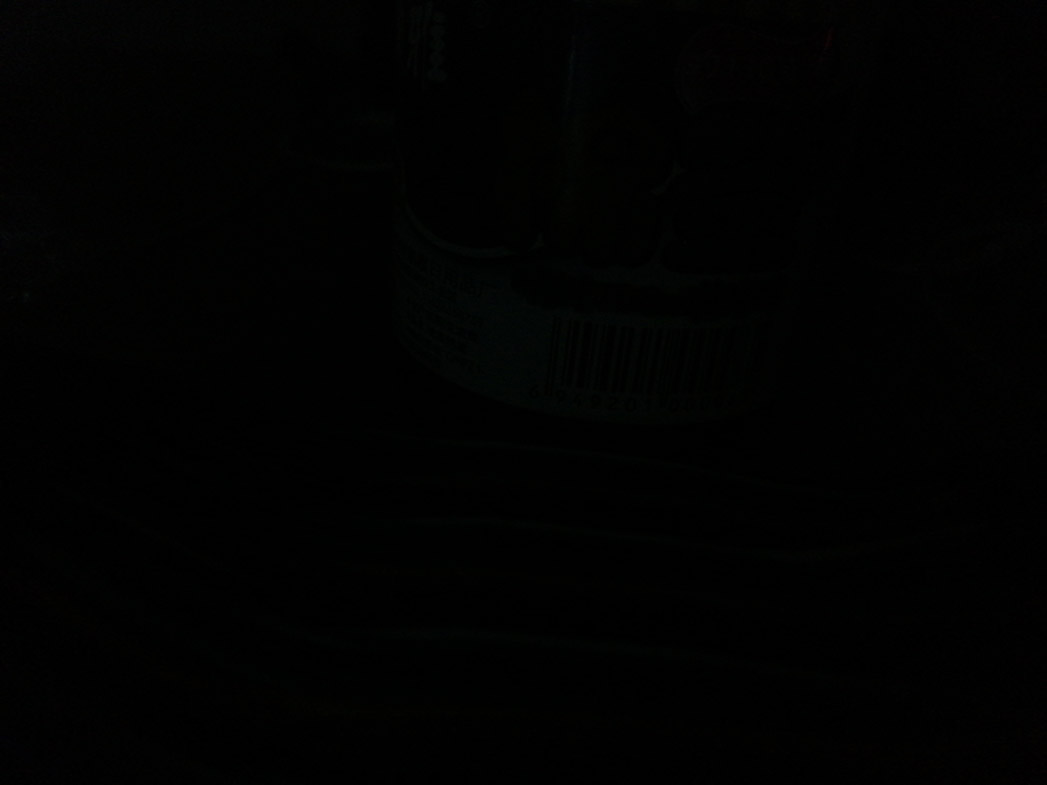}
    \end{minipage}   
  
    \\   
  
    \rotatebox[origin=c]{90}{our method}
    &
    \begin{minipage}{0.2\textwidth}
    \includegraphics[width=\textwidth]{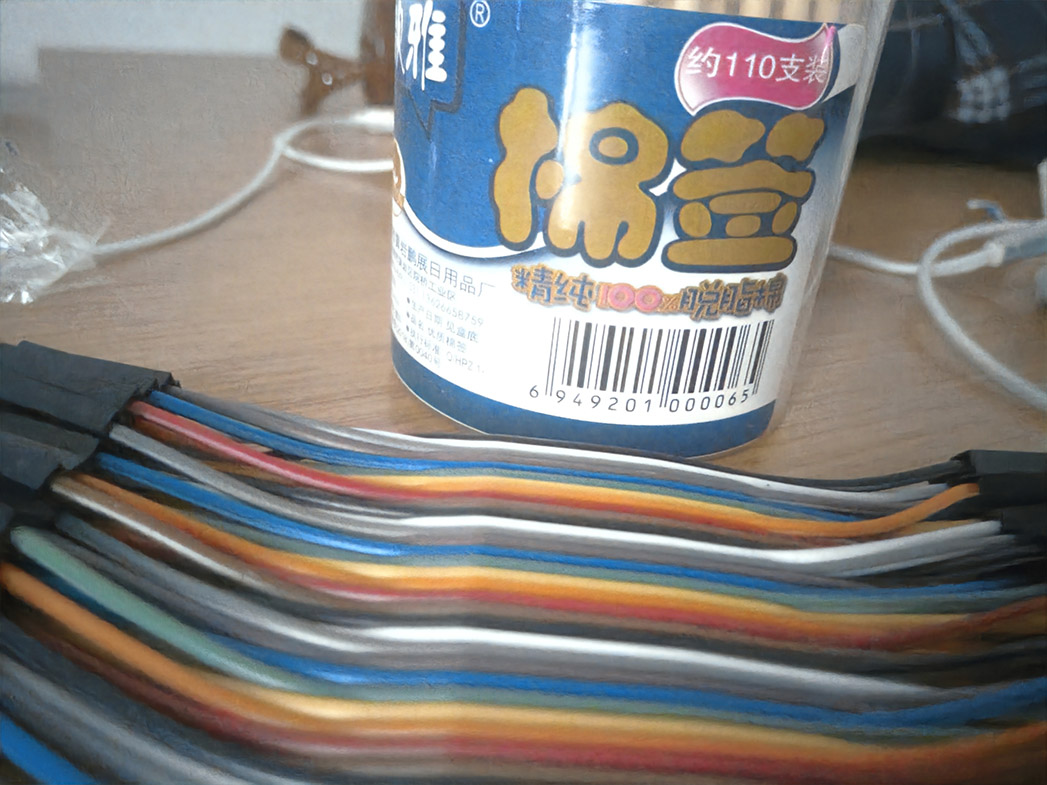}
    \end{minipage}   
    &   
    \begin{minipage}{0.2\textwidth}
    \includegraphics[width=\textwidth]{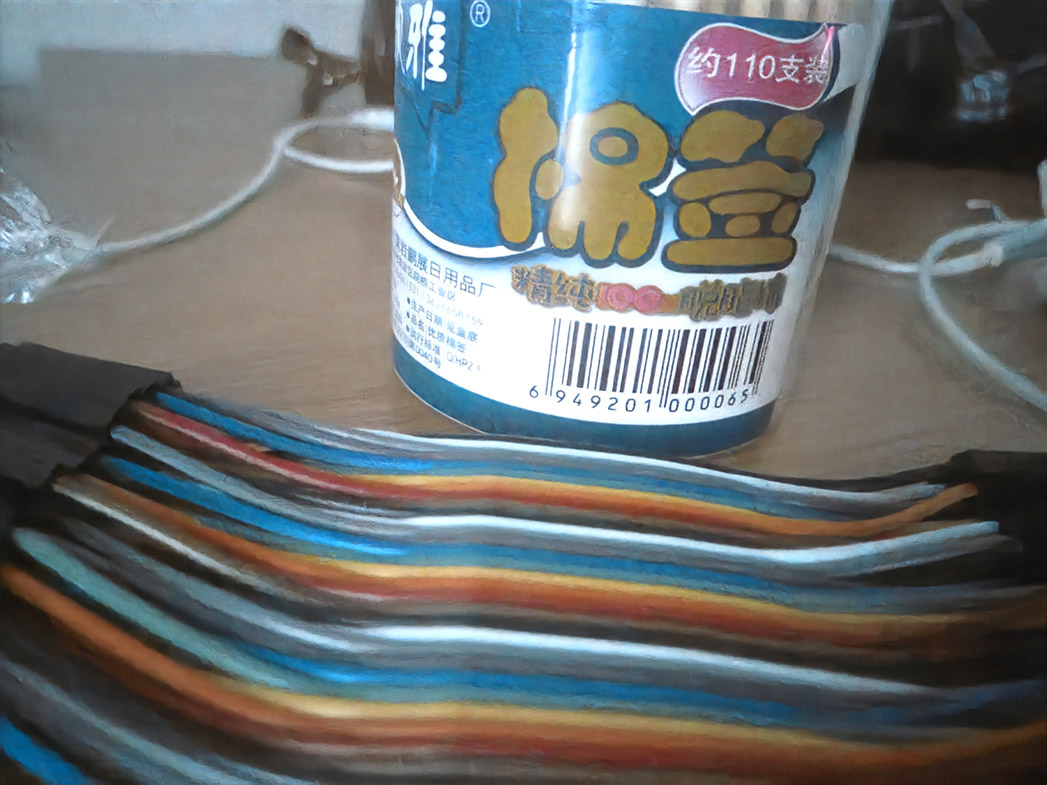}
    \end{minipage}   
    &   
    \begin{minipage}{0.2\textwidth}
    \includegraphics[width=\textwidth]{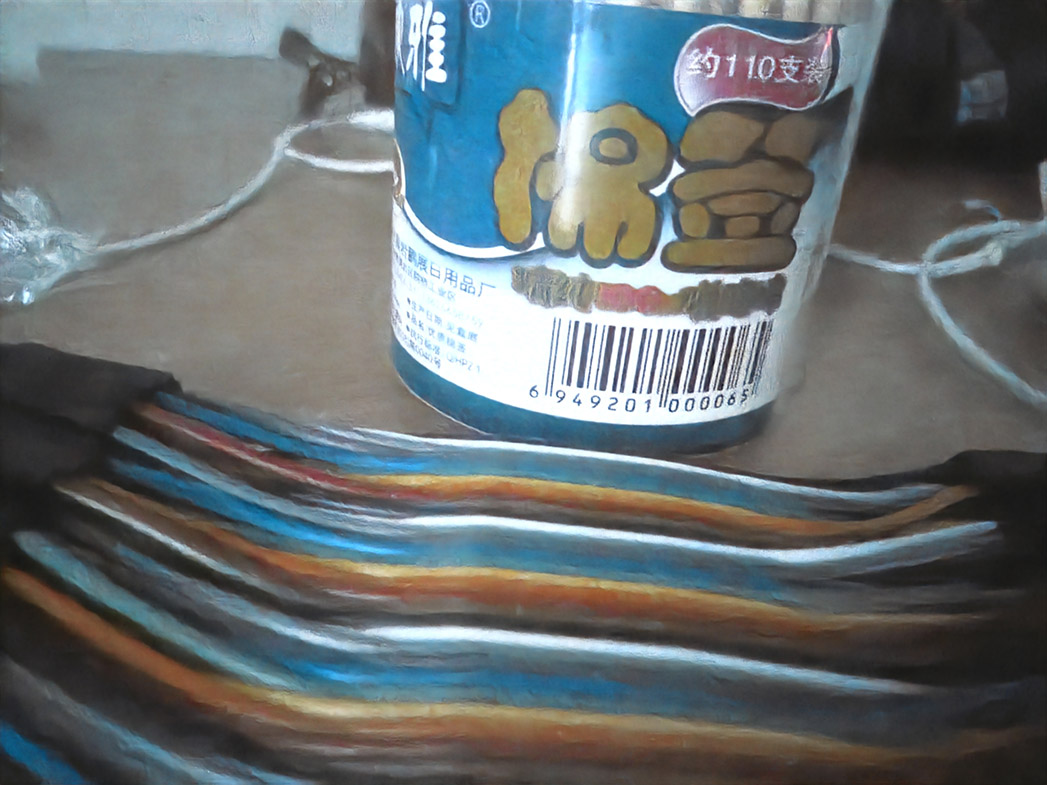}
    \end{minipage}   
    &   
    \begin{minipage}{0.2\textwidth}
    \includegraphics[width=\textwidth]{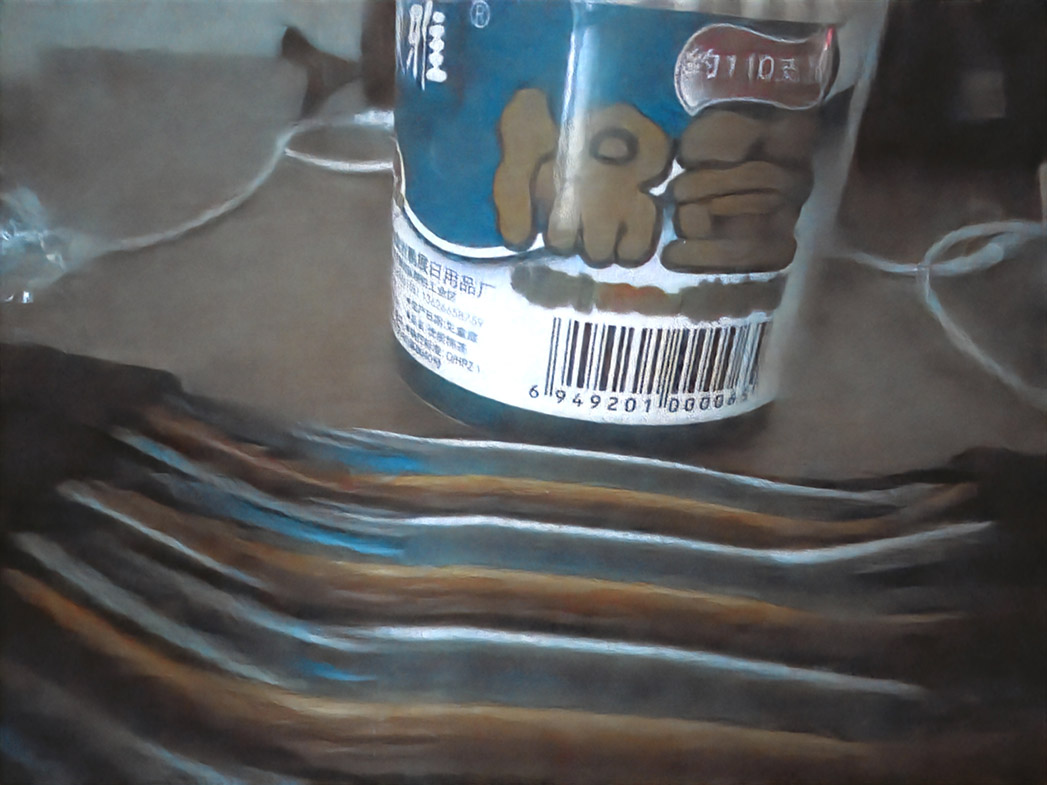}
    \end{minipage}   
    \\   

     (b) & 1/20 seconds & 1/60 seconds & 1/125 seconds  & 1/250 seconds  
  
  \end{tabular}
  \caption{The results of using our method to enhance multiple images 
  in a same scene but captured with diverse exposure time. 
  For the second image group, The exposure time $t_0$ of the last image in the sequence is only 8\% of that of the first, yet the network compensates for this difference, 
  keeping the brightness of each image in the scene approximately uniform.}
  \label{exp}
  \end{figure*}

\begin{figure*}[h!t] 
  \centering 
  \subfloat[original]{
    \begin{minipage}[b]{0.1292\linewidth} 
      \centering
      \includegraphics[width=\linewidth]{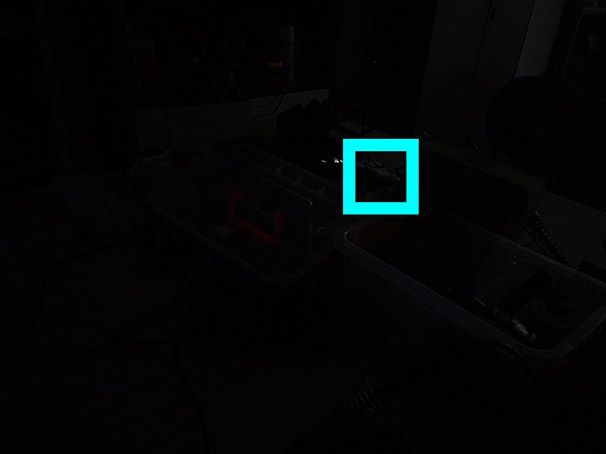}\vspace{4pt}
      \includegraphics[width=\linewidth]{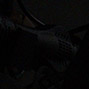}\vspace{4pt}
      \includegraphics[width=\linewidth]{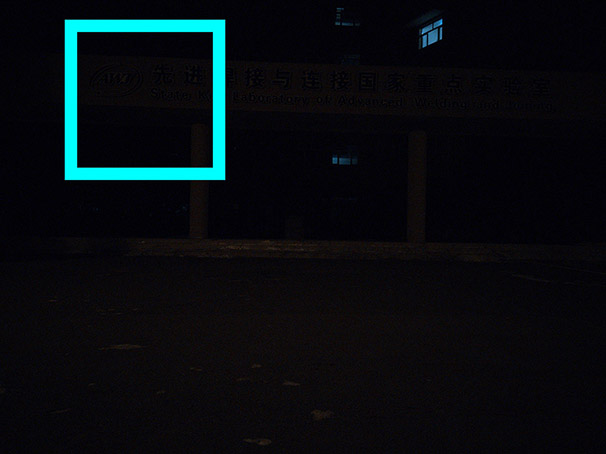}\vspace{4pt}
      \includegraphics[width=\linewidth]{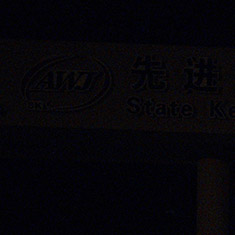}\vspace{4pt}
      \includegraphics[width=\linewidth]{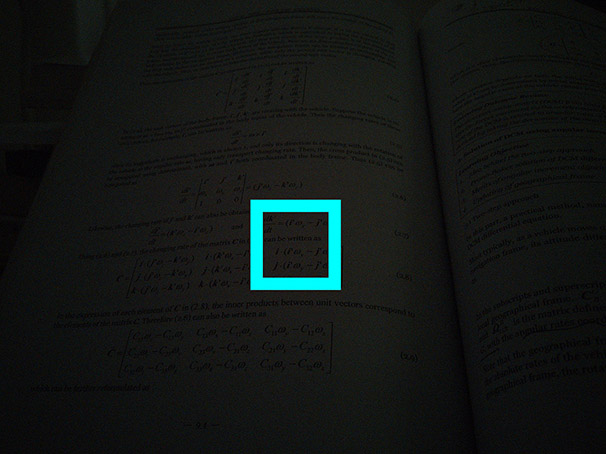}\vspace{4pt}
      \includegraphics[width=\linewidth]{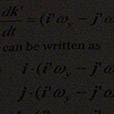}
    \end{minipage}
  }
  \hfill
    \subfloat[MSR]{
    \begin{minipage}[b]{0.1292\linewidth}
      \centering
      \includegraphics[width=\linewidth]{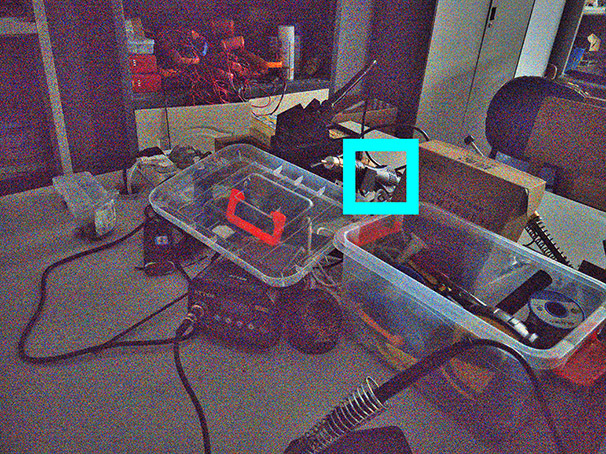}\vspace{4pt}
      \includegraphics[width=\linewidth]{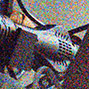}\vspace{4pt}
      \includegraphics[width=\linewidth]{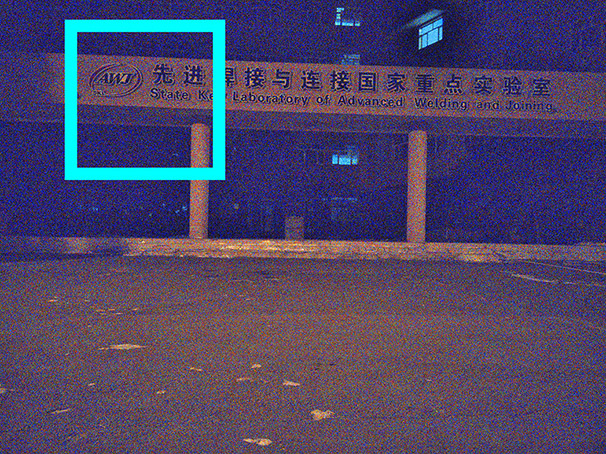}\vspace{4pt}
      \includegraphics[width=\linewidth]{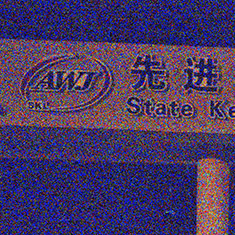}\vspace{4pt}
      \includegraphics[width=\linewidth]{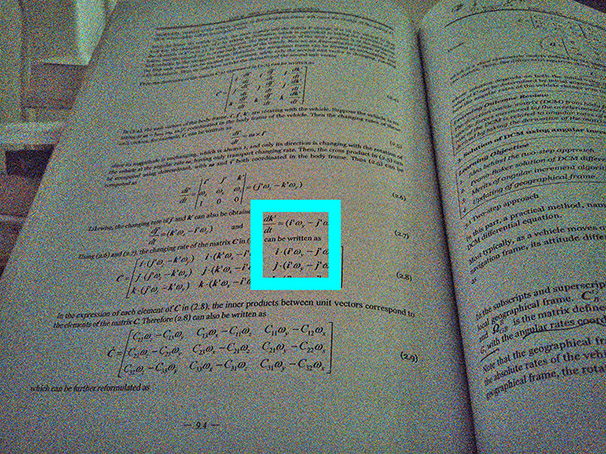}\vspace{4pt}
      \includegraphics[width=\linewidth]{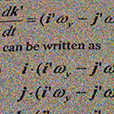}
    \end{minipage}
  }
  \hfill
    \subfloat[MSR+BM3D]{
    \begin{minipage}[b]{0.1292\linewidth}
      \centering
      \includegraphics[width=\linewidth]{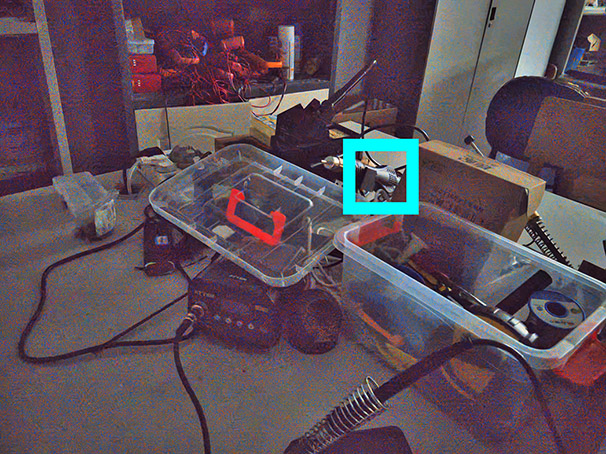}\vspace{4pt}
      \includegraphics[width=\linewidth]{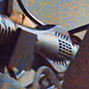}\vspace{4pt}
      \includegraphics[width=\linewidth]{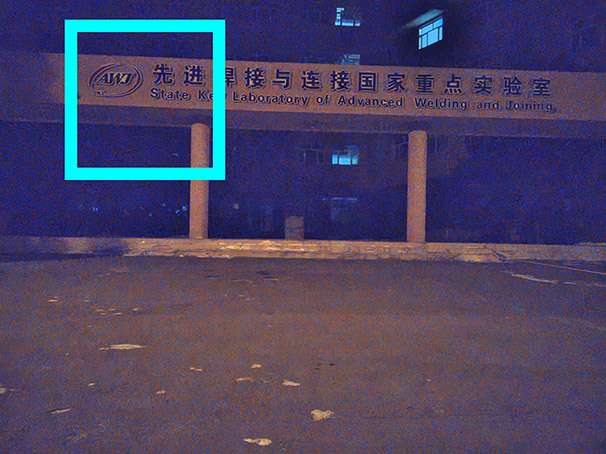}\vspace{4pt}
      \includegraphics[width=\linewidth]{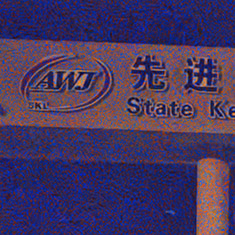}\vspace{4pt}      
      \includegraphics[width=\linewidth]{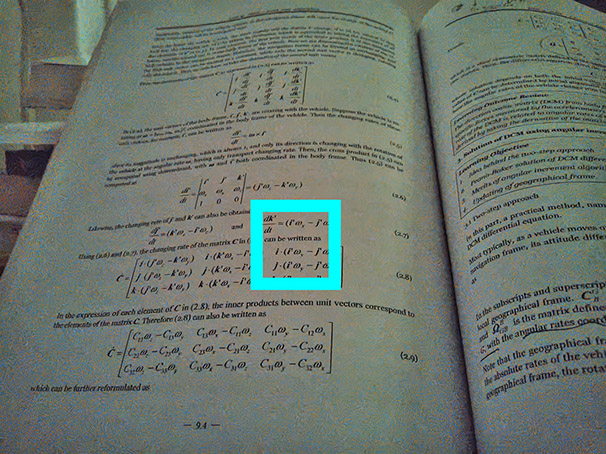}\vspace{4pt}
      \includegraphics[width=\linewidth]{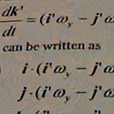}
    \end{minipage}
  }
  \hfill
    \subfloat[LIME+BM3D]{
    \begin{minipage}[b]{0.1292\linewidth}
      \centering
      \includegraphics[width=\linewidth]{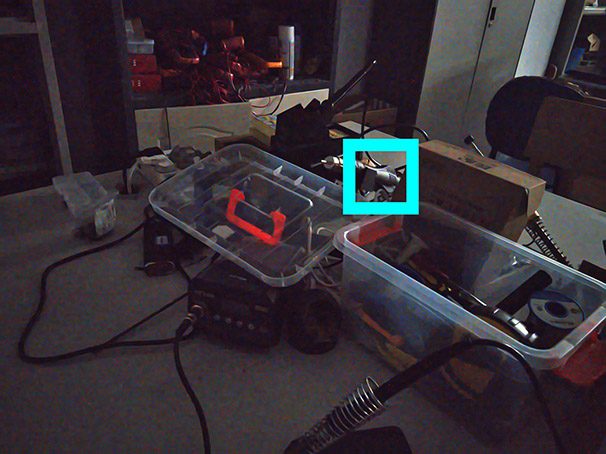}\vspace{4pt}
      \includegraphics[width=\linewidth]{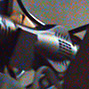}\vspace{4pt}
      \includegraphics[width=\linewidth]{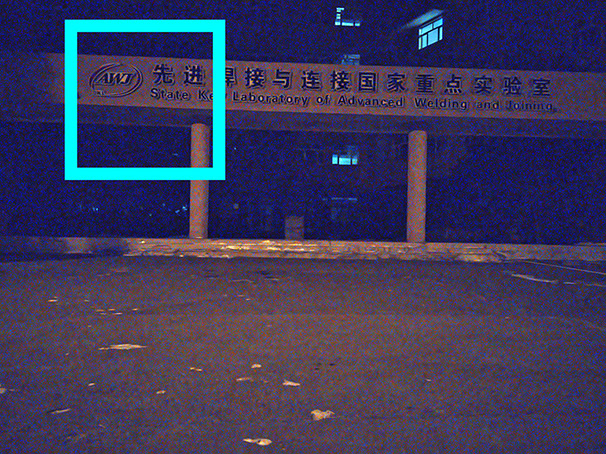}\vspace{4pt}
      \includegraphics[width=\linewidth]{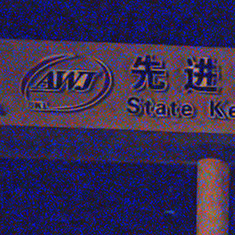}\vspace{4pt}      
      \includegraphics[width=\linewidth]{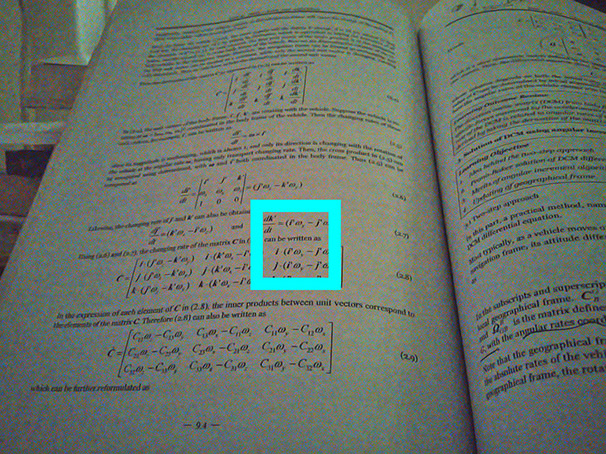}\vspace{4pt}
      \includegraphics[width=\linewidth]{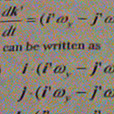}
    \end{minipage}
  }
  \hfill
    \subfloat[R-Net+BM3D]{
    \begin{minipage}[b]{0.1292\linewidth}
      \centering

      \includegraphics[width=\linewidth]{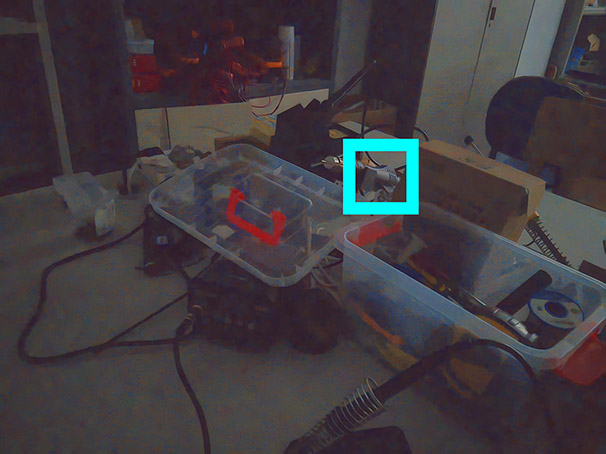}\vspace{4pt}
      \includegraphics[width=\linewidth]{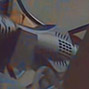}\vspace{4pt}
      \includegraphics[width=\linewidth]{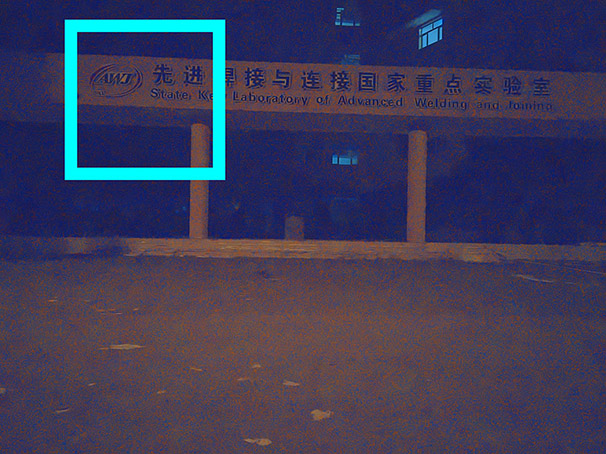}\vspace{4pt}
      \includegraphics[width=\linewidth]{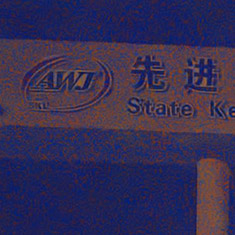}\vspace{4pt}      
      \includegraphics[width=\linewidth]{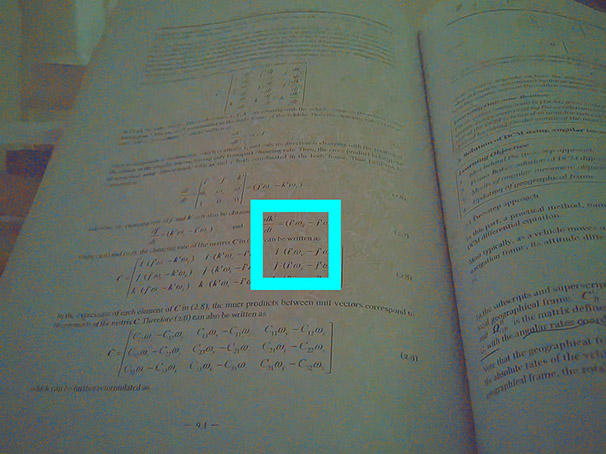}\vspace{4pt}
      \includegraphics[width=\linewidth]{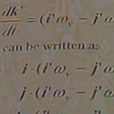}
    \end{minipage}
  }
  \hfill
  \subfloat[SID]{
  \begin{minipage}[b]{0.1292\linewidth}
    \centering

    \includegraphics[width=\linewidth]{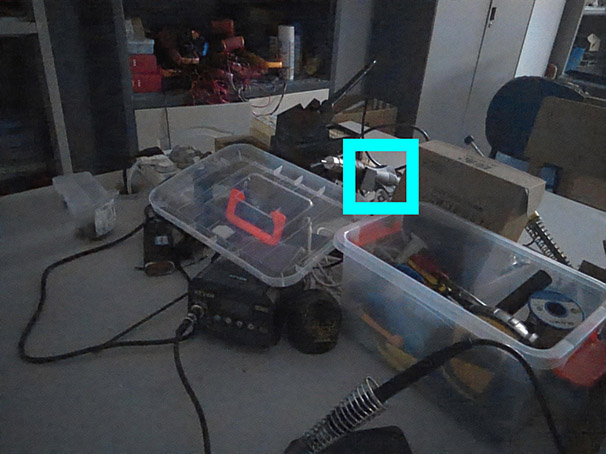}\vspace{4pt}
    \includegraphics[width=\linewidth]{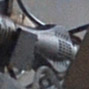}\vspace{4pt}
    \includegraphics[width=\linewidth]{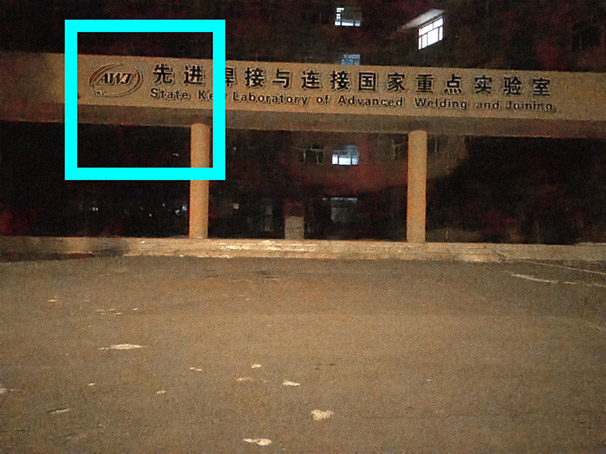}\vspace{4pt}
    \includegraphics[width=\linewidth]{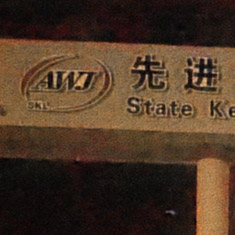}\vspace{4pt}    
    \includegraphics[width=\linewidth]{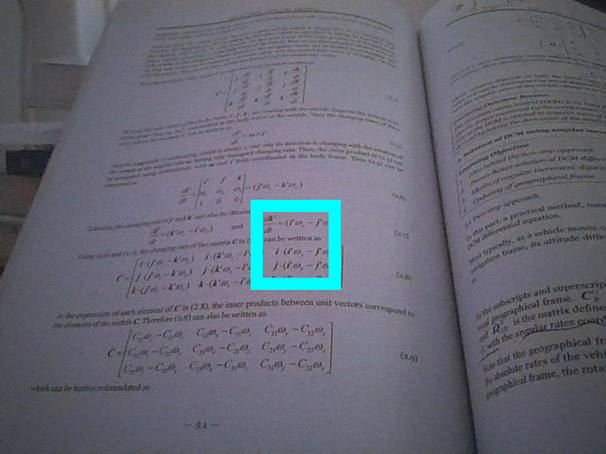}\vspace{4pt}
    \includegraphics[width=\linewidth]{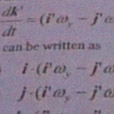}
  \end{minipage}
  }
  \hfill
  \subfloat[our method]{
  \begin{minipage}[b]{0.1292\linewidth}
    \centering
    \includegraphics[width=\linewidth]{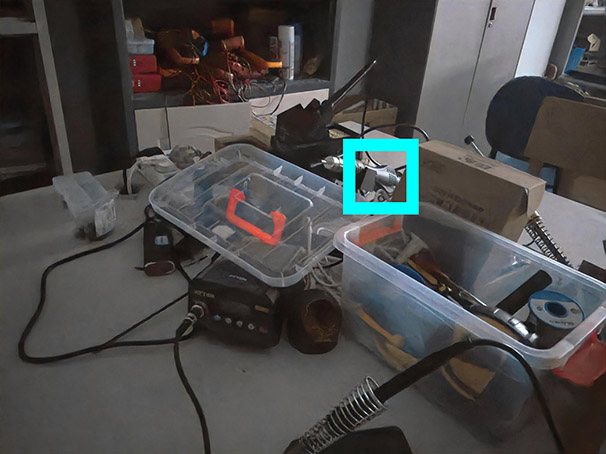}\vspace{4pt}
    \includegraphics[width=\linewidth]{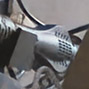}\vspace{4pt}
    \includegraphics[width=\linewidth]{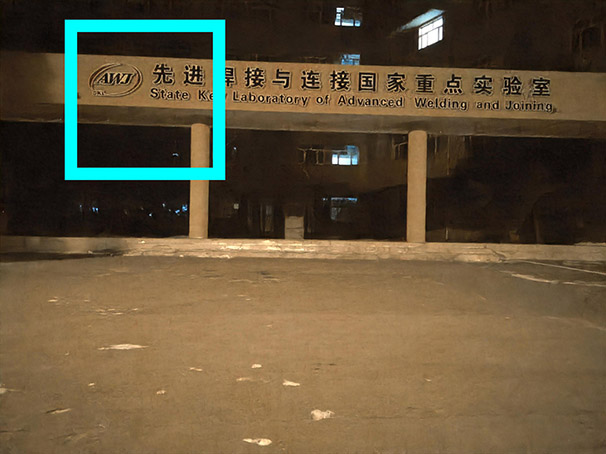}\vspace{4pt}
    \includegraphics[width=\linewidth]{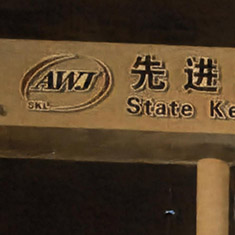}\vspace{4pt}
    \includegraphics[width=\linewidth]{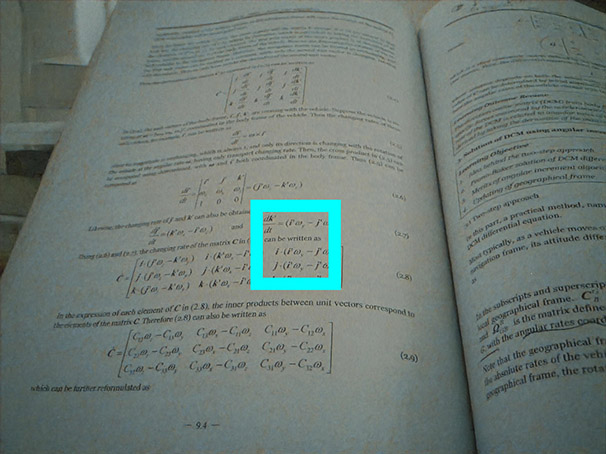}\vspace{4pt}
    \includegraphics[width=\linewidth]{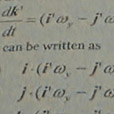}
  \end{minipage}
  }

  \caption{The results using different methods on CID test images.
  From (a) to (g), they are the original low-light images and the results of multiscale Retinex (MSR), MSR with BM3D,
  illumination map estimation based (LIME), LIME with BM3D, deep Retinex decomposition (R-net), R-net with BM3D, 
  learning-to-see-in-the-dark (SID) model and ours respectively.
  }
\label{compare}
\end{figure*}


The Brightness Prediction (BP) procedure provides the estimation of guideline exposure time $t_1$:
\begin{equation}
  \label{L_bpn1}
  t_1^{\Theta_2} = F_{BP}\left(\mathcal{X}_R,\mathcal{V}^{t_0}_p \middle| \Theta_2 \right)
\end{equation}
Where $F_{BP}$ and $\Theta_2$ is BPN and its parameters respectively. The pIEV is represented by $\mathcal{V}^{t_0}_p = \mathcal{V}_p(t_0)$.
It is the first enhancement procedure in our model but trained after the ESN.
As it is illustrated in Fig.\ref{bpn},
The Brightness Prediction Network (BPN) is a relatively small network, which consists of multiple convolution layers and then ends up with several fully connected layers.
Because of the existence of fully connected layers, 
the width and length of the input image $\mathcal{X}_R$ should be uniformly changed into $512 \times 512$.

With the estimated $t_1^{\Theta_2}$ obtained in this BP procedure, 
the $t_g$ item in IEV in the ES procedure can be replaced by $t_1^{\Theta_2}$.
Different from $\hat{\mathcal{Y}}^{\Theta_1}_{t_g}$ derived in Equation (\ref{L_esn}), the new final result image can be formulated as:
\begin{equation}
\begin{aligned}
  \label{L_esn2}
  &\hat{\mathcal{Y}}^{\Theta_2|\Theta_1^\ast} = F_{ES}\left(\mathcal{X}_{R},\mathcal{V}^{t_0 \rightarrow t_1^{\Theta_2}} \middle| \Theta_1^\ast \right)  \\
\end{aligned}
\end{equation}
Then an approach to train the BPN is required.
To begin with, the purpose of BPN is to control the brightness of the final result image,
which is achieved by adjusting the guiding exposure time $t_1$ in the ES procedure.
Since there are image pairs $\left(\mathcal{X}_R,\mathcal{Y}_R\right)$ and their exposure time $(t_0,t_g)$ available,
intuitively it seems that the BPN can simply be learned from $\left(\mathcal{X}_R,t_g\right)$ in an end-to-end way, with $\mathcal{X}_R$ as the input and $t_g$ as the label. 
However it cannot be achieved, the reason for which will be discussed later in the next subsection.

The proposed BPN training approach and the loss function $L_{BP}$ are as follows.
We define the brightness of a single pixel $Br$ as the average pixel value of all color channels.
Given a certain dark scene for image capturing and
with ISO, aperture settings and the scene all fixed, then the brightness of a pixel is only determined by exposure time.
Let $\mathcal{R}$ be the pixel irradiance in this scene, 
when the exposure time $t$ in camera settings increases, 
the pixel Exposure $\mathcal{E}(\mathcal{R},t)$ increases as well \cite{CRF_Estimation}:
\begin{equation}
  \label{L_Exposure}
  \mathcal{E}(\mathcal{R},t) = \mathcal{R}t
\end{equation}
Thus the pixel becomes brighter. 
It is feasible to design loss function $L_{BP}$ based on the pixel brightness.

It should be noted that the relationship between pixel brightness $Br(\mathcal{E})$ and Exposure $\mathcal{E}=\mathcal{R}t$ is not linear, 
instead, it can be typically described by an S-shaped curve because the pixel brightness is limited within $\left[0,1\right]$ 
(or $\left[0,255\right]$ for 8-bit image storage) 
and the camera sensor is less sensitive to the change of $t$ when the pixel value is close to $0$ or $1$.
In other words, as the exposure time $t$ changes, 
some pixels are not sensitive and there are little changes in the pixel brightness compared with some other pixels.
For example, when photographing a street at night, 
pixels revealing the lamp bulbs are always saturated unless the $t$ is extremely small,
and pixels representing the dark sky are always close to zero unless it is affected by noise.
If those pixels are involved in BPN training, they can cause gradient vanishing in the back propagation algorithm
because $\frac{\partial Br(\mathcal{E})}{\partial t} = \frac{\partial Br(\mathcal{E})}{ \partial \mathcal{E}} \cdot  \mathcal{R}$
and as it is revealed in the S-shaped culve, $\frac{\partial Br(\mathcal{E})}{ \partial \mathcal{E}}$ is close to 0.
On the other hand, 
when pixel brightness $Br$ is around $0.5$ the $\frac{\partial Br}{\partial t} $ reaches its maximum.
The pixels meeting this condition can be collected into an Area of Interest (AoI), 
which identifies pixels that are sensitive to the process of Exposure Shifting.

First, in the ground-truth image $\mathcal{Y}$, 
an AoI weight map $\mathcal{W}$ can be obtained by filtering out pixels with brightness near $0.5$.
More specifically, the ground-truth image $\mathcal{Y}$ is converted into a single channel grayscale image $\mathcal{Y}_G$,
then a Gaussian culve with mean $\mu_w = 0.5$ and variance $\sigma_w^2 = 0.01$ is applied on each pixel of $\mathcal{Y}_G$:
\begin{equation}
  \label{L_gaussian}
  \mathcal{W}_G = \exp{\left(-\frac{\left(\mathcal{Y}_G-\mu_w\right)^2}{2\sigma_w^2}\right)}
\end{equation}
Then $\mathcal{W}_G$ is normalized, the result is the AoI weight map $\mathcal{W}$. 
The value at position $(i,j)$ is denoted as subscripts, namely $\mathcal{W}_{G,ij}$ and $\mathcal{W}_{ij}$:
\begin{equation}
  \label{L_norm}
  \mathcal{W}_{ij}= \frac{ \mathcal{W}_{G,ij} }{ \sum_{i=1}^{m}\sum_{j=1}^{n}   \mathcal{W}_{G,ij}  },
  \forall 1 \leqslant i \leqslant m, 1 \leqslant j \leqslant n
\end{equation}
Combining the Equation \ref{L_gaussian} and \ref{L_norm}, they can be simplified by the softmax function:
\begin{equation}
  \label{L_weightmap}
  \mathcal{W}=softmax\left(-\frac{\left(\mathcal{Y}_G-\mu_w \right)^2}{2\sigma_w^2}\right)
\end{equation}
Lastly, the loss function $L_{BP}$ is designed to check if the final result image $\hat{\mathcal{Y}}^{\Theta_2|\Theta_1^\ast}$ has the same AoI.
Having the same AoI means that in this AoI area $\hat{\mathcal{Y}}^{\Theta_2|\Theta_1^\ast}$ and $\mathcal{Y}$ share similar brightness.
Moreover, recall that pixels outside AoI are not sensitive to the change of exposure time $t$ or the ES process,
thus it can be derived that the overall image brightness of $\hat{\mathcal{Y}}^{\Theta_2|\Theta_1^\ast}$ resembles that of $\mathcal{Y}$.
In this way, BPN completed the adaptive Brightness Prediction objective.

Then the enhanced image $\hat{\mathcal{Y}}^{\Theta_2|\Theta_1^\ast}$ is converted into a grayscale image $\hat{\mathcal{Y}}^{\Theta_2|\Theta_1^\ast}_G$.
When the AoI of $\hat{\mathcal{Y}}^{\Theta_2|\Theta_1^\ast}$ and $\mathcal{Y}$ overlaps exactly, the following loss function reaches its minimum:
\begin{equation}
\begin{aligned}
  \label{L_lossbp}
  L_{BP} &= L_{BP}\left(\hat{\mathcal{Y}}^{\Theta_2|\Theta_1^\ast}_G,\mathcal{W}  \middle|  \Theta_2;\Theta_1^\ast       \right) \\
  &= -\frac{1}{mn}\sum_{i=1}^{m}\sum_{j=1}^{n}{\exp{\left(-\frac{\left(   \hat{\mathcal{Y}}^{^{\Theta_2|\Theta_1^\ast}}_{G,ij}-\mu_w\right)^2}{2\sigma_v^2}\right)}\mathcal{W}_{ij}}
\end{aligned}
\end{equation}
Where $\sigma_v^2 = 0.04$ is another variance constant. The image value at position $(i,j)$ is represented by $i,j$ subscripts. Finally, BPN is trained by optimizing $\Theta_2$ over $K$ pairs of images in the training set:
\begin{equation}
  \label{L_f}
  \Theta_2^\ast=\mathop{\argmin}_{\Theta_2} \sum_{k=1}^{K} L_{BP}\left(\hat{\mathcal{Y}}^{\Theta_2|\Theta_1^\ast}_{G,k},\mathcal{W}_k  \middle|  \Theta_2;\Theta_1^\ast       \right)
\end{equation}

\subsection{Method Summary and Discussion}

In Algorithm 1, the training process of our model is summarized, 
together with the method to evaluate and apply this model.
\begin{algorithm}
\label{A1}
\caption{Model training and evaluation}  
\begin{algorithmic}[1] 
\State Initialize ESN, BPN parameters $\Theta_1$ and $\Theta_2$
\State Prepare CID dataset
\Function {Train}{Dataset}
\For{$EpochTrainESN = 1 \to E_1$}  
  \For{Image Pair $ k = 1 \to K$}   
  \State Read raw image pairs $\left(\mathcal{X}_{R,k},\mathcal{Y}_{R,k}\right)$ and $\mathcal{V}^{t_0 \rightarrow t_g}_k$
  \State Convert to RGB format. $\mathcal{Y}_{k} \gets \mathcal{Y}_{R,k}$
  \State Compute ${\hat{\mathcal{Y}}}^{\Theta_1}_k$ via Equation \ref{L_esn}
  \State $\Theta_1 \gets \mathop{\argmin}_{\Theta_1} {\sum_{k=1}^{K}{L_{ES} \left(  {\hat{\mathcal{Y}}}^{\Theta_1}_k , \mathcal{Y}_k    \right)    }}$
  \EndFor 
\EndFor 
\State $\Theta_1^\ast \gets \Theta_1$

\For{$EpochTrainBPN = 1 \to E_2$}  
  \For{Image Pair $ k = 1 \to K$}
  \State Read raw image pair $\left(\mathcal{X}_{R,k},\mathcal{Y}_{R,k}\right)$ and $\mathcal{V}^{t_0}_{p,k}$
  \State Compute $t_{1,k}^{\Theta_2}$ and $\hat{\mathcal{Y}}^{\Theta_2|\Theta_1^\ast}_k$ via Equation \ref{L_bpn1} and \ref{L_esn2}
\State Convert to grayscale image. $\mathcal{Y}_{G,k} \gets \mathcal{Y}_{R,k}$ 
\State Convert to grayscale image. $\hat{\mathcal{Y}}^{\Theta_2|\Theta_1^\ast}_{G,k} \gets \hat{\mathcal{Y}}^{\Theta_2|\Theta_1^\ast}_k$
  \State  Compute $\mathcal{W}_k$ via Equation \ref{L_weightmap}
  \State $\Theta_2 \gets \mathop{\argmin}_{\Theta_2} \sum_{k=1}^{K} L_{BP}\left(\hat{\mathcal{Y}}^{\Theta_2|\Theta_1^\ast}_{G,k},\mathcal{W}_k  \middle|  \Theta_2;\Theta_1^\ast       \right)$
  \EndFor 
\EndFor 
\State $\Theta_2^\ast \gets \Theta_2$
\State \Return{$\Theta_1^\ast , \Theta_2^\ast$}
\EndFunction
\\
\Function {Evaluate}{Image}
\State Read raw-format image $\mathcal{X}_{R}$
\State Read all elements in $\mathcal{V}^{t_0}_{p}$ from raw-image file
\State BPN: Compute $t_1^{\Theta_2^\ast} \gets F_{BP}\left(\mathcal{X}_R,\mathcal{V}^{t_0}_p \middle| \Theta_2^\ast \right)$
\State ESN: Compute $\hat{\mathcal{Y}}^{\Theta_2^\ast|\Theta_1^\ast} \gets F_{ES}\left(\mathcal{X}_{R},\mathcal{V}^{t_0 \rightarrow t_1^{\Theta_2^\ast}} \middle| \Theta_1^\ast \right)$
\State \Return $\hat{\mathcal{Y}}^{\Theta_2^\ast|\Theta_1^\ast}$
\EndFunction
\end{algorithmic}
\end{algorithm}

As mentioned earlier in the last section, 
training BPN with input $\mathcal{X}_R$ and label $t_g$ is straightforward but not feasible.
First let $F_{BP}^\ast$ be the hypothetical optimal BPN model, 
then $t_1^\ast(\mathcal{X}_R) = F_{BP}^\ast(\mathcal{X}_R)$ is the optimal estimation of guideline exposure time.
For a pair of image and time label $\left(\mathcal{X}_R,t_g\right)$,
the $t_g$ can be considered as one sampling from a Gaussian distribution: $t_g \sim \mathcal{N}\left(t_1^\ast,\sigma_c^2 \right)$.
The variance $\sigma_c^2$ in this process is too large, the reason are:
\begin{itemize}
\item In every CID image group, the ground-truth image is chosen manually and is dominated by subjective human judgments.
\item The ground-truth image is selected from only 8 candidate images, but the optimal $t_1^\ast$ exists somewhere in the time continuum $\left(0,\inf \right)$.
\end{itemize}
Furthermore, the number of samples is very limited since each image group can only provide one $t_g$ sample.
As a result, this straightforward approach cannot be applied in the training of BPN due to the high variance of the label $t_g$ and the limited number of samples.

\section{Experiments}\label{sec5}
\subsection{Training}

Limited by the GPU memory, we did not adopt the Batch Normalization \cite{ioffe2015batch} technique.
To accelerate model training, before the training process started,
each channel of the input image, as well as each element of IEV, was normalized along the sample dimension.
Then the means and variances used in the normalization became invariant constants.
Additionally, images for training were randomly cropped into $512\times512$ patches. 

Both ESN and BPN were trained with Adam optimizer\cite{kingma2014adam}, following the procedures shown in Algorithm 1. 
The ESN was trained first with the learning rate slowly descending from $2 \times 10^{-4}$ to $1 \times 10^{-5}$. 
After about 300 epochs, when the loss on training sets became stable, the parameters in ESN were made untrainable. 
Then BPN was appended to the training graph and trained with the same learning rate. 
Considering BPN may require global information to decide the proper exposure time, 
we used bigger patches but avoid training with the whole image to prevent overfitting. 
Although the loss must propagate through the deep ESN before reaching BPN, 
the training process went smoothly and completed after 100 epochs.

\subsection{Content-Based Brightness Control}

\begin{figure*}[h]
  \centering
  \begin{tabular}{ccc}

    \begin{minipage}{0.2\textwidth}
    \includegraphics[width=\textwidth]{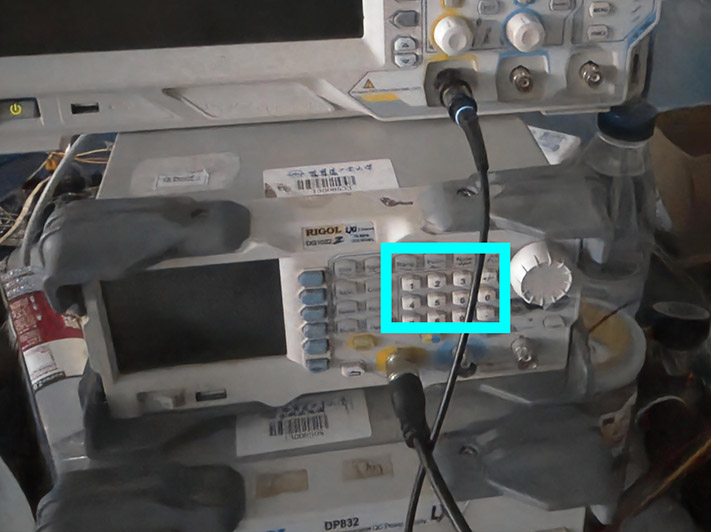}\vspace{4pt}
    \end{minipage}   
    &   
    \begin{minipage}{0.2\textwidth}
    \includegraphics[width=\textwidth]{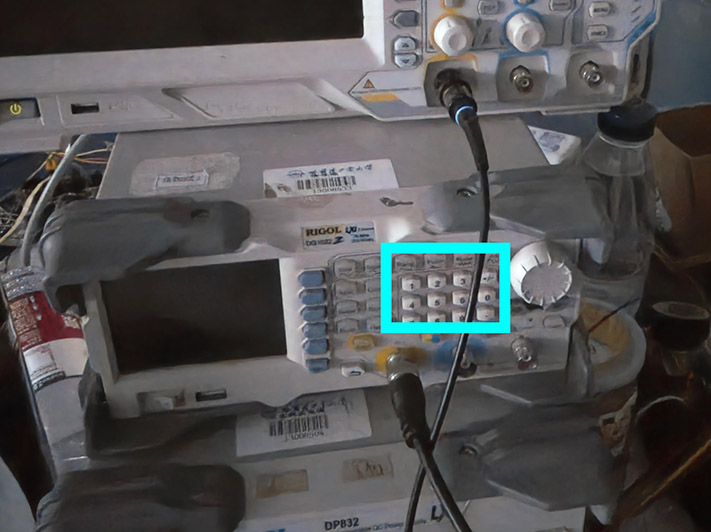}\vspace{4pt}
    \end{minipage}   
    &   
    \begin{minipage}{0.2\textwidth}
    \includegraphics[width=\textwidth]{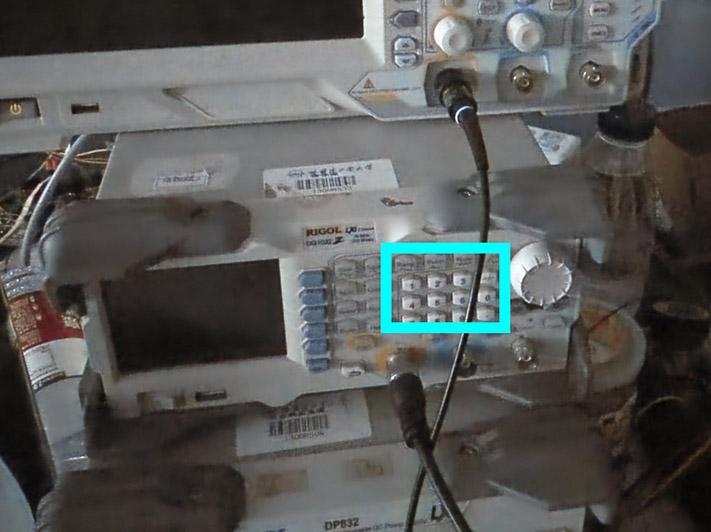}\vspace{4pt}
    \end{minipage}   

    \\   
    \begin{minipage}{0.2\textwidth}
    \includegraphics[width=\textwidth]{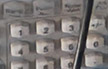}
    \end{minipage}   
    &   
    \begin{minipage}{0.2\textwidth}
    \includegraphics[width=\textwidth]{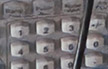}
    \end{minipage}   
    &   
    \begin{minipage}{0.2\textwidth}
    \includegraphics[width=\textwidth]{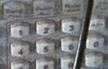}
    \end{minipage}   

    \\   
    
    $\alpha=0$  &  $\alpha=0.15$    &   $\alpha=0.30$
  
    \\

    $L_1=L_{MAE}$ &   $L_1=0.85L_{MAE}+0.15L_{SSIM}$    &  $L_1=0.7L_{MAE}+0.3L_{SSIM}$

  \end{tabular}
  \caption{Controlled experiments on the impact of SSIM loss.}
  \label{ssim}
  \end{figure*}

\begin{figure}[!t]
    \centering
    \includegraphics[width=2.2in]{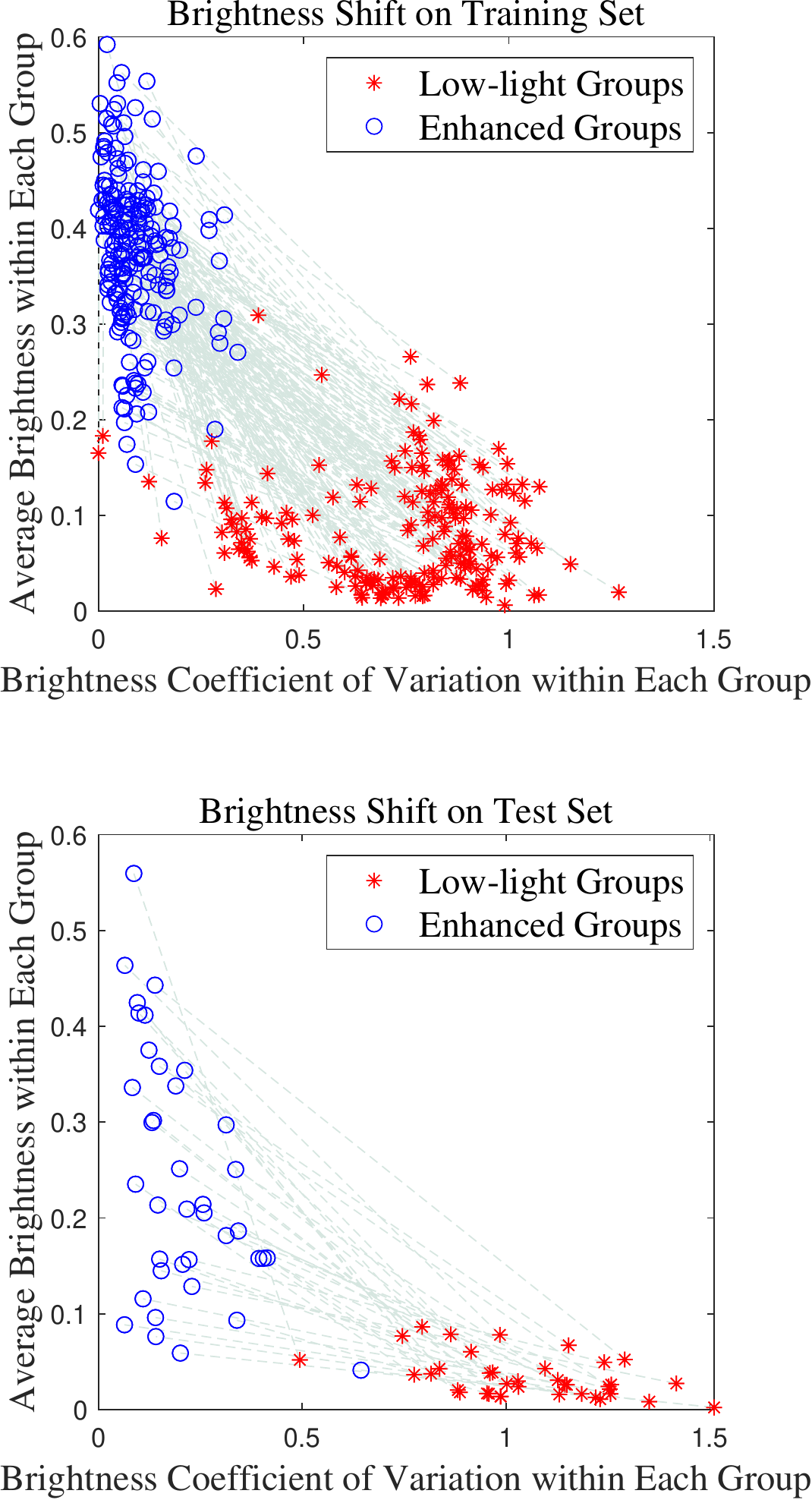}
    \caption{The shift of average brightness and brightness coefficient of variation (CV) 
    within each group. Each Group contain 3--6 images captured in a brust with different 
    exposure time, e.g. Fig.\ref{exp}. In both test set and training set, an increase in brightness 
    and reduction in CV is observed. Note that the CV decreased considerably, indicating the 
    highly uniformed brightness for images within each group after enhancement.}
    \label{brightnessshift}
    \end{figure}
  \begin{figure}[!t]
    \centering
    \includegraphics[width=2.5in]{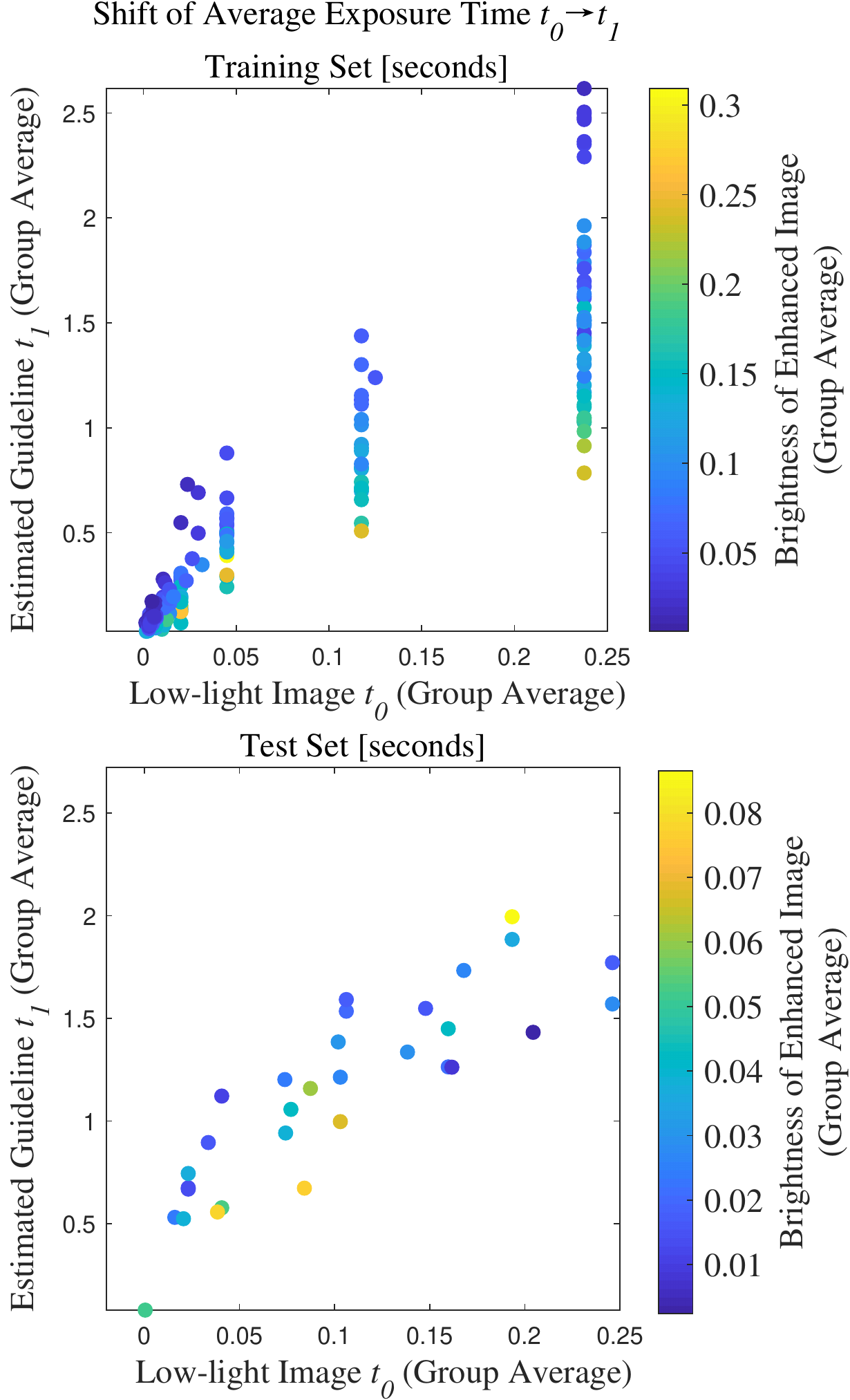}
    \caption{The comparison of image original exposure time $t_0$ and BPN assigned time $t_1$. 
    BPN infers $t_1$ based on not only the original time $t_0$ but also the content of the scene. 
    The color indicates the mean global brightness value of enhanced images within each group.}
    \label{timeshift}
    \end{figure}

\begin{table}[!t]
    \processtable{Comparison results with non-reference image quality metrics.
    MSR, LIME, R-Net methods are combined with BM3D to make a fair comparison.
    No-reference quality metrics, including Statistical Noise Level Estimation (SNLE), NLE(Noise Level Estimation), NV(Noise Variance) and image entropy, 
    are applied to evaluate model denoising and brightness control performance.\label{nrm}}
    {\begin{tabular*}{20pc}{@{\extracolsep{\fill}}ccccc@{}}\toprule
        Methods               & SNLE  & NLE & NV & Entropy \\  
    \midrule
    Original              & -                   & -                 &-                          & 4.0689 \\  
    MSR+BM3D              & \textbf{0.0014}     & 1.3792            &0.8998                     & 7.0068 \\  
    LIME+BM3D             & 0.0086              & 4.2437            &10.1789                    & \textbf{7.1084} \\  
    R-Net+BM3D            & 0.0017              & 0.2467            &\textbf{0.4491}            & 6.1503 \\  
    Ours                  & \textbf{0.0015}     & \textbf{0.1061}   &\textbf{0.4554}            & \textbf{6.6112} \\
    \botrule
    \end{tabular*}}{}
    \end{table}

\begin{figure}[htb]
  \centering
  \includegraphics[width=3.5in]{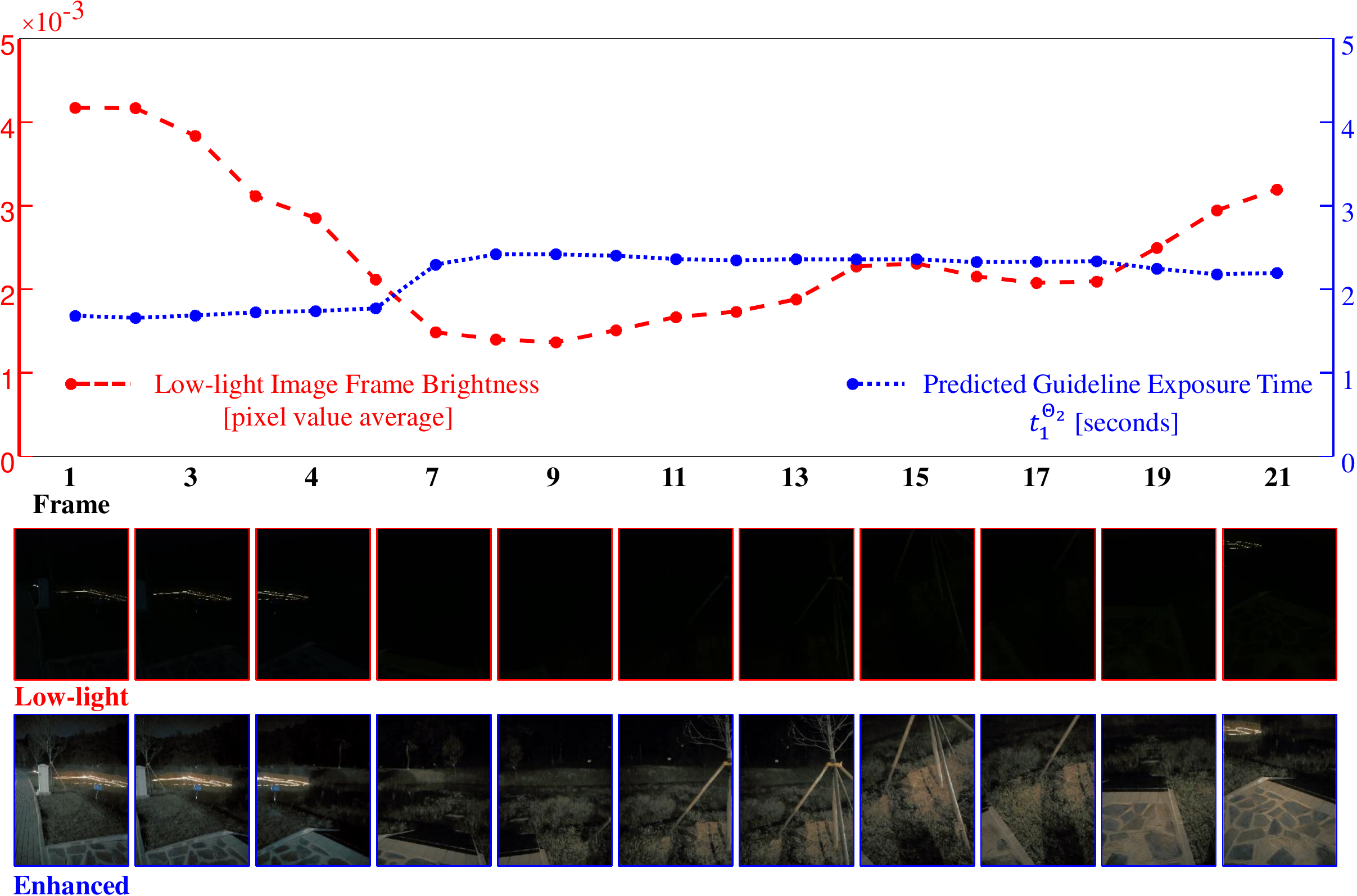}
  \caption{Applying our method to dynamic low-light video enhancement.}
  \label{video}
  \end{figure}
Firstly, we highlight the adaptive brightness control feature of our method. 
Here we explicitly define the brightness as the mean value of an image $\mathcal{Y}$, denoted as $Br(\mathcal{Y})$.
We expect a dynamic adjustment of the image brightness based on the image content. 
More specifically, 
while extreme low-light images suffer from severe noise along with color distortion and are barely visible before post-processing,
other mild low-light images only have moderate noise and relatively lower brightness compared with normal images.
Thus, the enhancement algorithm needs to make adjustments adaptively for different input images.
Currently the classic methods, 
such as algorithms based on Retinex and Dehazing, 
have poor performance on the extreme low-light images,
On the other hand, 
learning-based methods targeting at end-to-end denoising can process all kinds of low-light images, 
but they lack the flexibility and adaptability 
because they either need brightness amplification hand-tuning, 
or simply serve as a denoising procedure in other low-light methods.  
The proposed model has both the advantages and is able to process low-light images of different levels.

In order to evaluate the adaptability, we apply the trained model to enhance all images in the test set.
Recall that the image group is the basic unit of our CID dataset, 
accordingly, instead of evaluating single images, 
the collaborative characteristics across different images are investigated within each image group.
Note that the group identity of an image is NOT involved in model training 
and only serves as a tag to gather resulting images for further analysis.

For a sequence of multi-exposed images $(\mathcal{X}_{R,1},\mathcal{X}_{R,2},...)$ with increasing exposure time,
it is expected that after enhancement the resulting images possess similar image brightness, 
namely $Br(\hat{\mathcal{Y}}_1) \approx Br(\hat{\mathcal{Y}}_2)\approx ...$, 
where $\hat{\mathcal{Y}}$ denotes the enhanced image.
Fig.\ref{exp}(a) illustrates a busy-road scene, the first row displays the RGB images produced by the camera,
the second displays the enhanced results by utilizing our method on individual raw images. 
Note that the first image in this group is exposed for 1/20 seconds, 
causing motion blur to the truck, while the last image is exposed for only 1/100 seconds. 
It is shown that all four images have approximately uniform brightness after the image enhancement, 
despite the changing $t_0$ and the moving vehicle headlight.
The performance of our method on extreme low-light conditions is shown in Fig.\ref{exp}(b). 
The last image in the sequence suffers from both low environment illumination and short exposure time and the color and shape of objects in the image are drowned out by noise. However, 
the resulting image has the noise reduced and still maintain acceptable global brightness.

As mentioned above, for quantitative analysis, 
we gathered images that belong to the same groups and calculated the brightness of all images in order to observe the shift of brightness.
Let $(Br_1,Br_2,...)_k$ be the image brightness of each unenhanced low-light image in $k$-th image group
and let $(Br'_1,Br'_2,...)_k$ be that of the enhanced ones.
Then,
the mean value and the CV (coefficient of variation) of $(Br_1,Br_2,...)$ is computed and denoted as $(\mu_,c_v)_k$.
Correspondingly, $(\mu',c_v')_k$ is calculated from $(Br'_1,Br'_2,...)_k$.
In Fig.\ref{brightnessshift}, 
we plotted $(\mu_,c_v)$ and $(\mu',c_v')$ for all groups.
To illustrate the change after enhancement, 
for each group $(\mu_,c_v)$ (marked with star) and $(\mu',c_v')$ (marked with circle) are linked with a dotted line.
It can be observed from Fig.\ref{brightnessshift} that for different groups that captured from diverse scenes, 
the enhanced images all end up with increased image brightness and significantly reduced CV.
And yet the brightness growth among the groups can be very different, ranging from around 60\% to over 1,000\%. 
Considering together with Fig.\ref{exp}, 
Fig.\ref{brightnessshift} indicates that the BP sub-model is able to adaptively control the global image brightness 
not only based on the brightness of the original image, but also according to the content of the scene.
On the other hand, the dramatic reduction of CV suggests that the enhanced images within each group share almost identical brightness in spite of the varying exposure time $t_0$. 
To be precise, 
our model can control group brightness CV under 0.29 on 76.3\% test set groups and 98.5\% training set groups
considering the influence of the diverse scenes, illumination conditions, ISO etc.

When designing our model, 
we expect the BPN to extract features of the scene content and then estimates a reasonable $t_1$ accordingly. 
In order to see through clearly how this black box works, 
the relationships among the real exposure time $t_0$, 
the BPN-estimated guideline exposure time $t_1$
and the brightness of enhanced images $Br(\hat{\mathcal{Y}})$ are shown in Fig.\ref{timeshift}.
Similar to Fig.\ref{brightnessshift}, we use group average to simplify the figure, in which each point corresponds to an image group. 
As shown in Fig.\ref{timeshift}, the network tends to estimate $t_1$ positively associated with $t_0$. 
Furthermore, for groups with identical average $t_0$, 
the contents of the scenes have an impact on $t_1$ prediction. 
To be specific, if the scene of a group has more relatively bright areas, 
a smaller $t_1$ is more likely to be adopted to prevent overexposure and vice versa.
And yet a few exceptions exist, 
suggesting the BPN sub-model has learned some more sophisticated rules in the neural network black box.

\subsection{Denoising Evaluation}

We compare our method with four classic and state-of-art methods, 
including multiscale Retinex (MSR)\cite{jobson1997multiscale}, illumination map estimation based (LIME)\cite{guo2017lime}, 
deep Retinex decomposition (Retinex-Net, R-Net)\cite{wei2018deep} and learning-to-see-in-the-dark (SID)\cite{chen2018learning}. 
For fair comparison, 3D transform-domain filtering (BM3D)\cite{Kostadin2007Image} is applied on MSR, 
LIME, R-Net, since those methods do not model noise and require extra denoising process. Additionally, we train the SID model on the CID dataset and manually select ratio $\gamma$ for the SID model since it does not provide automatic brightness estimation.

Recall that the BPN utilizes a reduced reference evaluation index as the loss in its training.
In this way, the exposure parameters of the ground-truth image, which is chosen by highly subjective human judgment, 
will have much less influence on BPN model training.
As a result of this novel technique, however, PSNR and SSIM evaluation cannot be adopted because there is no reference image. 
We provide perceptual comparisons in Fig.\ref{compare} and evaluate our method with no-reference quality metrics.

In Table.\ref{nrm}, we compare the proposed method with MSR, LIME and R-Net(Retinex-Net),
using several no-reference quality metrics including Statistical Noise Level Estimation (SNLE) \cite{Chen2015AnSNLE},
Noise Level Estimation (NLE) \cite{liu2013single}, Noise Variance (NV) \cite{immerkaer1996fast} and image entropy.
SNLE, NLE and NV metrics can estimate noise level from a single image, 
and clean images have relatively smaller values than the noisy ones.
The SNLE estimates noise variance by observing the eigenvalues of images 
and the NLE by applying the principal component analysis (PCA) on special image patches.
The NV metric is motivated by the fact that clean and noisy images have different sensitivity to the Laplacian operation.
The image entropy reflects the average amount of information in an image, 
it is the most simple image evaluation metric. 
Images with proper brightness have more visible information, thus have larger image entropy compared to under-exposed or over-exposed ones.
We run evaluations on about 400 CID test images and the SID method is excluded 
because it requires an external manually-selected amplification ratio for every image.
As it is shown in Table.\ref{nrm}, our method has superior performance on Noise Level Estimation.
Moreover, the no-reference metrics use very different standards to assess image quality.
MSR+BM3D, R-Net+BM3D and LIME+BM3D methods all have undesired scores on one or multiple indices,
while our method has good performance on all metrics.

It is illustrated in Fig.\ref{compare} that our method and SID method provide more significant improvements in noise removal than the classic method BM3D and can adaptively denoise images with diverse noise levels. 
Furthermore, our results benefit from the white balance index introduced in IEV and avoid abnormal color in extreme low-light images.

However, the CNN-based method tends to blur the object with sharp edges (e.g. text). 
By introducing SSIM into loss function, our method shows more promising results in images with text 
and has advantages over SID in keeping more edge information, e.g. the images in the first row of Fig.\ref{compare}.
But on the other hand, in controlled experiments shown in Fig.\ref{ssim}, it is revealed that the ratio of SSIM component $\alpha$ in loss function $L_{ES}$ 
has to be contained to avoid the noise being interpreted into texture and edge by ESN.

\subsection{Low-light Video Processing}
The advantage of our work lies in getting rid of the brightness control ratio while achieving state-of-the-art low-light noise suppression. 
Without any handcrafting parameters as network input, we can directly apply our method 
to raw low-light video processing.

A raw-format video can occupy a very large amount of storage space. 
Limited by devices, 
we only developed a relatively simple way to verify our method on raw low-light videos.
In Fig.\ref{video}, 
we experiment on 21 continuous raw images captured in a burst session, 
shot with short exposure time and identical ISO, 
to discuss the potential applications in raw video processing. 
In this scene, the camera starts from a bridge with lighting decoration, 
then turns to the dark riverbank, 
and finally to the dimly illuminated sidewalk. 
As is shown in Fig.\ref{video}, 
though the enhancement process of each frame is completely independent of other frames, 
neither the estimated $t_1$ nor the resulting images show any sudden fluctuation. 
However, for practical considerations, it is also convenient to implement guideline time filters between ESN and BPN, 
which allows the value of $t_1$ to be dynamically filtered or amplified according to different demands.

\subsection{Computational Cost}

The proposed algorithm is evaluated on the ModelArts cloud service, 
which is equipped with 8 vCPUs and an Nvidia-P100 GPU (16GB GPU memory).
The execution time for a high-resolution image ($3968 \times 2976$) is $0.27$ seconds on average, 
which only shows a minor increase compared with the SID algorithm ($0.21$ seconds).
More specifically, the execution time of BPN and ESN sub-model is $0.05$ and $0.22$ respectively.
For comparison, it takes $4.32$ seconds for BM3D (GPU implementation \cite{bm3d-gpu}) to complete the denoising process, 
therefore other low-light methods that depend on BM3D for denoising, 
including MSR, LIME, R-Net et al., 
are much less efficient on devices with GPUs.

\section{Conclusion and Discussion}\label{sec6}

\subsection{Conclusion}
In this paper, an adaptive raw image enhancement model is proposed for extreme low-light image processing. 
The two-stage framework provides adaptability to images with different scenes and exposure parameters. 
We also presented the CID dataset for model training and evaluation. 
The experimental test shows that our model can provide the state-of-the-art low-light image denoising as well as adaptive global brightness control. 

The proposed method has many advantages over existing approaches.
In the first place, no external parameters are needed after model training, 
a single raw image and its Exif metadata (e.g. white balance, ISO and exposure time) are sufficient to complete the process. 
Therefore, our model is able to process a large batch of raw images without parameter-handcrafting.
Secondly, our model significantly outperforms other methods in which the denoising stage works as a separable step.
Because instead of simply implementing noise-suppression algorithms (e.g. BM3D) before or after the main low-light enhancement procedure,
we make denoising procedure an integrated part in our model, 
allowing the ESN module to learn exposure shifting and denoising simultaneously in an end-to-end way.
In quantitative experiments, 
we adopted no-reference image quality metrics including SNLE, NLE, NV and image entropy to test the denoising performance.
Our model has the smallest noise variance on the NLE metric compared to BM3D-based low-light methods 
and obtains near-optimal scores on the other three indices.

We also reveal the potential application in low-light video processing. 
The BPN module makes it possible for our model to process images with diverse exposure levels, 
from extreme low-light images to mild ones. 
But to be applied in video processing, it should face more challenges, 
e.g. the brightness of frames should change continuously and avoid fluctuation. 
We experiment on continuous raw image series, despite that no information of adjacent frames is used, 
the output of BPN changes continuously with the illumination condition and shows no fluctuation.

\subsection{Limitations and Future Works}

The model proposed is limited by dataset, which is a common problem in data-driven methods.
On the one hand, the model is trained by raw-format and multi-exposed images,
which significantly improved image quality.
But on the other hand, all reference ground-truth images are captured in the real scenes, 
as a result, they sometimes have minor defects such as noise, halos around the light source, local over-exposure and white balance deviation.
The denoising performance may be further improved if those drawbacks can be avoided.
Then the image groups in CID have not covered enough nature scenes.
For example, it is found that green grass and trees tend to have low color saturation in the resulting images, 
that is because most images in the training set are collected in winter when there is hardly any green plants in scenes. 
For future study, 
the raw-format CID dataset can be extended by unprocessing \cite{brooks2019unprocessing} technique and generating raw-format images from other available online datasets.

Another limitation is the memory consumption problem. 
The batch size is constrained because it takes too much GPU memory to train the proposed model.
Therefore training the model can be time-consuming on GPUs with small memory.
We are planning to improve network architecture and investigate the possibility of application on mobile devices.

As a future expectation, the model can be further modified by introducing HDRI (High Dynamic Range Imaging).
Since our CID dataset is composed of multi-exposed images, 
we can calculate an HDR image of each scene.
Finally, the HDR images can be used as ground truth to participate in the ESN training.

\bibliographystyle{iet}
\bibliography{sample_sim}

\end{document}